\documentclass[twocolumn]{aastex631}

\usepackage{amsmath,amsthm,amsfonts,amssymb,amscd,bm}
\usepackage{booktabs}

\begin{document}
\title{Accurate Measurement of the Lensing Magnification by BOSS CMASS Galaxies and Its Implications for Cosmology and Dark Matter}

\correspondingauthor{Kun Xu, Y.P. Jing}
\email{woshixukun@sjtu.edu.cn,
 ypjing@sjtu.edu.cn}

\author[0000-0002-7697-3306]{Kun Xu}
\affil{Department of Astronomy, School of Physics and Astronomy, Shanghai Jiao Tong University, Shanghai, 200240, People’s Republic of China}
\affil{Institute for Computational Cosmology, Department of Physics, Durham University, South Road, Durham DH1 3LE, UK}

\author[0000-0002-4534-3125]{Y.P. Jing}
\affil{Department of Astronomy, School of Physics and Astronomy, Shanghai Jiao Tong University, Shanghai, 200240, People’s Republic of China}
\affil{Tsung-Dao Lee Institute, and Shanghai Key Laboratory for Particle Physics and Cosmology, Shanghai Jiao Tong University, Shanghai, 200240, People’s Republic of China}

\author{Hongyu Gao}
\affil{Department of Astronomy, School of Physics and Astronomy, Shanghai Jiao Tong University, Shanghai, 200240, People’s Republic of China}

\author[0000-0002-4574-4551]{Xiaolin Luo}
\affil{Department of Astronomy, School of Physics and Astronomy, Shanghai Jiao Tong University, Shanghai, 200240, People’s Republic of China}

\author[0000-0002-1318-4828]{Ming Li}
\affil{National Astronomical Observatories, Chinese Academy of Sciences, Beijing, 100101, People’s Republic of China}



\begin{abstract}
Magnification serves as an independent and complementary gravitational lensing measurement to shear. We develop a novel method to achieve an accurate and robust magnification measurement around BOSS CMASS galaxies across physical scales of $0.016h^{-1}{\rm Mpc} < r_{\rm p} < 10h^{-1}{\rm Mpc}$. We first measure  the excess total flux density $\delta M$ of the source galaxies in deep DECaLS photometric catalog that are lensed by CMASS galaxies. We convert $\delta M$ to magnification $\mu$ by establishing the $\delta \mu-\delta M$ relation using a deeper photometric sample. By comparing magnification measurements in three optical bands ($grz$), we constrain the dust attenuation curve and its radial distribution, discovering a steep attenuation curve in the circumgalactic medium of CMASS galaxies. We further compare dust-corrected magnification measurements to model predictions from high-resolution dark matter-only (DMO) simulations in WMAP and Planck cosmologies, as well as the hydrodynamic simulation \texttt{TNG300-1}, using precise galaxy-halo connections from the Photometric objects Around Cosmic webs method and the accurate ray-tracing algorithm \texttt{P3MLens}. For $r_{\rm p} > 70h^{-1}$ kpc, our magnification measurements are in good agreement with both WMAP and Planck cosmologies, resulting in an estimation of the matter fluctuation amplitude of $S_8=0.816\pm0.024$. However, at $r_{\rm p} < 70h^{-1}$ kpc, we observe an excess magnification signal, which is  higher than the DMO model in Planck cosmology at $2.8\sigma$ and would be exacerbated if significant baryon feedback is included.  Implications of the potential small scale discrepancy for the nature of dark matter and for the processes governing galaxy formation are discussed.
\end{abstract}

\keywords{Gravitational lensing (670) --- Dark matter (353) --- Observational cosmology (1146) --- Intergalactic dust clouds (810)}


\section{Introduction} \label{sec:intro}
Light from distant sources undergoes gravitational bending as it traverses through the Universe due to the gravitational potential of foreground large-scale structures, a phenomenon known as gravitational lensing \citep{1993ApJ...404..441K,2001PhR...340..291B}. Gravitational lensing holds significant potential as a cosmological probe, offering a direct measure of the distribution of matter. Gravitational lensing manifests in two main effects: shear and magnification. Shear describes the shape distortion of sources behind a lens. Magnification refers to the amplification of the surface area behind a lens, resulting in the magnification of the flux of background sources, as lensing conserves the apparent surface brightness.

Most existing weak lensing measurements in galaxy surveys primarily utilize shear \citep{2019A&A...625A...2K,2022PhRvD.105b3520A,2023PhRvD.108l3521S}, as the signal of magnification is considerably weaker at cosmological scales \citep{2000A&A...353...41S}. However, unbiased shear measurements encounter numerous challenges, including the accurate measurement of galaxy shapes \citep{2018ARA&A..56..393M} and contamination from intrinsic alignments \citep{2000ApJ...545..561C,2004PhRvD..70f3526H,2009ApJ...694..214O,2023ApJ...954....2X}. Magnification, on the other hand, can utilize fainter and unresolved sources. Moreover, at small scales, where the signals for both shear and magnification are sufficiently strong, the noise associated with shear may exceed that of magnification, making magnification a more effective probe at these scales. Consequently, magnification is emerging as a complementary probe to shear, providing independent measurements of the matter distribution.

To date, most studies have treated magnification as a source of contamination in galaxy clustering and shear measurements \citep{2008PhRvD..78l3517Z,2020A&A...636A..95D,2021JCAP...12..009M,2021MNRAS.504.1452V,2024MNRAS.527.1760W}, because magnification alters the observed spatial distribution of galaxies, a phenomenon known as `magnification bias'. Direct measurements of magnification are typically conducted using the cross-correlation of foreground and background samples \citep{1995ApJ...438...49B,1998ApJ...501..539T,2005ApJ...633..589S,2010MNRAS.405.1025M,2010MNRAS.406.2352L,2011MNRAS.414..596W,2013MNRAS.429.3230H,2014MNRAS.440.3701B,2014MNRAS.442.2680G,2019JCAP...09..021B,2024A&A...684A.109C}, similar to galaxy-galaxy lensing in shear measurements. This is necessary because the intrinsic clustering of galaxies introduces significant fluctuations in number density, which greatly exceed those due to lensing effects. \citet{2005ApJ...633..589S} achieved the first measurements of magnification on cosmological scales using large samples of quasars (QSOs) as sources and galaxies as lenses from the Sloan Digital Sky Survey \citep[SDSS;][]{2000AJ....120.1579Y}. \citet{2010MNRAS.405.1025M} obtained improved measurements using larger samples. \citet{2014MNRAS.440.3701B} attempted to measure magnification using optical galaxy samples as sources, but the signal was much weaker due to the low redshift of the sample. High-redshift galaxy tracers, such as sub-millimetre galaxies \citep{2010MNRAS.406.2352L,2011MNRAS.414..596W,2014MNRAS.442.2680G,2019JCAP...09..021B,2024A&A...684A.109C} and Lyman-break galaxies \citep{2013MNRAS.429.3230H}, are considered promising source samples for measuring magnification due to their high redshift and steep number counts. Although many attempts have been made to measure magnification, its application in cosmology is limited due to both the insufficient accuracy of measurements and the immaturity of models \citep{2020A&A...639A.128B,2021A&A...646A.152G}.

Furthermore, dust attenuation within the large-scale structure can contaminate magnification measurements \citep{2010MNRAS.405.1025M,2013MNRAS.429.3230H,2014MNRAS.440.3701B}. Dust in the lenses can absorb light from background soueces, altering their apparent flux. This alteration is degenerate with magnification effects caused by lensing. However, since magnification effects are wavelength-independent while dust attenuation depends on wavelength, this challenge becomes an opportunity to constrain dust attenuation in the large-scale structure by comparing magnification measurements across different wavelengths \citep{2010MNRAS.405.1025M}. By correcting for dust attenuation, the true magnification signal can be extracted. Consequently, both the mass and dust distributions can be simultaneously constrained in magnification measurements.

In this work, we introduce a novel method to measure magnification by measuring the change of total flux density \(\delta M\) of sources around lens galaxies and converting it to the lens parameter \(\mu\). This conversion is achieved by establishing the \(\delta \mu-\delta M\) relation using a deeper photometric survey. With this method, we successfully measure the magnification signal around Baryon Oscillation Spectroscopic Survey (BOSS) CMASS galaxies using source galaxies from the deep Dark Energy Camera Legacy Survey (DECaLS). We obtain robust magnification measurements at physical scales of \(0.016h^{-1}{\rm Mpc} < r_{\rm p} < 10h^{-1}{\rm Mpc}\). By comparing magnification measurements in the \(grz\) bands, we constrain the dust attenuation curve and its radial distribution in the circumgalactic medium (CGM) of CMASS galaxies, deriving the true magnification signal after correcting for dust attenuation. We find a steep dust attenuation curve in the CGM of CMASS galaxies.

Next, we compare our observations to high-resolution cosmological simulations using the precise galaxy-halo connection of CMASS from the Photometric Objects Around Cosmic Webs (PAC) method \citep{2023ApJ...944..200X} and the accurate ray-tracing algorithm \texttt{P3MLens} \citep{2021ApJ...915...75X}. We evaluate our findings against dark matter-only (DMO) simulations under the nine-year Wilkinson Microwave Anisotropy Probe (WMAP9) and Planck18 cosmologies, as well as the hydrodynamic simulation \texttt{TNG300-1}. At \(r_{\rm p} > 70h^{-1}\) kpc, our magnification measurements align with both WMAP9 and Planck18 cosmologies. However, we observe an excess magnification signal at \(r_{\rm p} < 70h^{-1}\) kpc compared to DMO simulations, and including the hydrodynamic model only increases the discrepancy. Our findings may indicate an incomplete understanding of the nature of dark matter or the physics of galaxy formation. Additionally, our study demonstrates that magnification is a promising probe for constraining galaxy formation and cosmology.

We present the methodology in Section \ref{sec:method}. The observation data is detailed in Section \ref{sec:obs_data}. Measurements of magnification and constraints on dust attenuation are discussed in Sections \ref{sec:measurements} and \ref{sec:dust}, respectively. The simulation data is introduced in Section \ref{sec:simulation}, and the model predictions are provided in Section \ref{sec:model_pred}. Finally, the discussion and conclusion are given in Sections \ref{sec:discussion} and \ref{sec:conclusion}.

\section{Lensing magnification and its observables}\label{sec:method}
In this section, we provide a brief explanation of the magnification effect in gravitational lensing, and discuss the observables employed for its measurement in galaxy surveys.

\subsection{Magnification Effect in Gravitational Lensing}\label{sec:basics}
Assuming that gravitational lensing alters the source position from $\bm{\theta}^S$ to $\bm{\theta}^I$, this transformation can be expressed by the following matrix:
\begin{equation}
    \mathbf{A} = \frac{\partial \bm{\theta}^S}{\partial \bm{\theta}^I} =
    \begin{pmatrix}
        1-\kappa-\gamma_1 & -\gamma_2 \\
        -\gamma_2 & 1-\kappa+\gamma_1
    \end{pmatrix}\,,
\end{equation}
where $\kappa$ represents the convergence, and $\gamma_1$ and $\gamma_2$ denote the two components of the shear, with ${\bm{\gamma}} = \gamma_1 + i\gamma_2$. Magnification characterizes the amplification of the surface area:
\begin{equation}
\mu = \left| \frac{\partial \bm{\theta}^I}{\partial \bm{\theta}^S} \right| = \frac{1}{(1-\kappa)^2-\gamma^2}.\label{eq:mu}
\end{equation}
As lensing conserves the apparent surface brightness, the total flux of the background sources is magnified by $\mu$.

The lensing parameters are linked to the matter distribution in the lens plane. The convergence $\kappa$ is linked to the matter surface density $\Sigma$ through the critical surface density $\Sigma_{\rm{crit}}$:
\begin{equation}
\kappa(\bm{\theta})=\frac{\Sigma(\bm{\theta})}{\Sigma_{\rm{crit}}}\,,\ \Sigma_{\rm{crit}}=\frac{c^2}{4\pi G}\frac{D_{\rm{A}}^{\rm{s}}}{D_{\rm{A}}^{\rm{l}}D_{\rm{A}}^{\rm{ls}}}\,,
\end{equation}
where $D_{\rm{A}}^{\rm{l}}$, $D_{\rm{A}}^{\rm{s}}$, and $D_{\rm{A}}^{\rm{ls}}$ are the angular diameter distances from the observer to the lens, the observer to the source, and from the lens to the source. For a lens with an {\textit{axially symmetric}} mass distribution, the shear amplitude $\gamma$ can be computed as:\
\begin{equation}
\gamma(\theta)=\frac{\Delta \Sigma(\theta)}{\Sigma_{\rm{crit}}}\,,\ \Delta \Sigma(\theta)=\bar{\Sigma}(\theta)-\Sigma(\theta)\,,
\end{equation}
where $\bar{\Sigma}(\theta)$ represents the mean matter surface density enclosed within $\theta$:
\begin{equation}
\bar{\Sigma}(\theta)=\frac{1}{\pi\theta^{2}}\int_0^{\theta}d\theta^{'}2\pi\theta^{'}\Sigma(\theta^{'})\,.
\end{equation}
In the weak lensing regime (\(\kappa \ll 1\)), magnification can be approximated as:
\begin{equation}
\mu \approx 1 + 2\kappa\,.
\end{equation}
Thus,
\begin{equation}
\delta\mu(\theta) = \mu(\theta) - 1 \approx \frac{2\Sigma(\bm{\theta})}{\Sigma_{\rm{crit}}}\,.\label{eq:delmu}
\end{equation}
However, in regions of high density where the nonlinear contribution cannot be ignored, \(\delta\mu\) may significantly deviate from the approximation in Equation \ref{eq:delmu}. Therefore, we define:
\begin{equation}
    \Sigma_{\mu} = \frac{\delta\mu \Sigma_{\rm{crit}}}{2}\,,
\end{equation}
and \(\Sigma_{\mu} \approx \Sigma\) in the weak lensing regime. \(\Sigma_{\mu}\) is defined to extract the redshift dependence and facilitate the comparison of magnification measurements across different source samples. For source samples with different redshift distributions, \(\Sigma_{\mu}\) is theoretically different due to its remaining dependence on source redshifts. However, for source samples with similar redshifts, as used in this study, the difference is negligible since the redshift dependence of \(\Sigma_{\mu}\) is very weak.

% as per Equation \ref{eq:mu}, isolating the redshift dependence in the $\mu-\Sigma$ relation is challenging. However, if the source samples have very close redshifts, as in this study, the majority of the redshift dependence can be extracted by defining
% \begin{equation}
%     \Sigma_{\mu} = \frac{\delta\mu\Sigma_{\rm{crit}}}{2}\,,
% \end{equation}
% and $\Sigma_{\mu}\approx\Sigma$ in the weak lensing regime. As demonstrated in Section \ref{sec:simu_lens}, we find that the remaining redshift dependence is negligible ($<3\%$) for our source samples.

In this study, we investigate lensing magnification in the vicinity of massive galaxies, where the matter density can be notably high and asymmetric. Consequently, we opt for direct ray-tracing in cosmological simulations to compare with observations \citep{2000ApJ...530..547J,2016A&C....17...73P,2021ApJ...915...75X}, rather than relying on analytical models with assumptions. After solving for the deflection angle $\bm{\alpha}$ in the lens plane, the lensing parameters can be derived by:
\begin{equation}
    \gamma_1 = \frac{D_{\rm{A}}^{\rm{l}}D_{\rm{A}}^{\rm{ls}}}{2D_{\rm{A}}^{\rm{s}}}\left(\frac{\partial\alpha_x}{\partial x}-\frac{\partial\alpha_y}{\partial y}\right)\,,\notag
\end{equation}
\begin{equation}
    \gamma_2 = \frac{D_{\rm{A}}^{\rm{l}}D_{\rm{A}}^{\rm{ls}}}{2D_{\rm{A}}^{\rm{s}}}\left(\frac{\partial\alpha_x}{\partial y}+\frac{\partial\alpha_y}{\partial x}\right)\,,\notag
\end{equation}
\begin{equation}
    \kappa = \frac{D_{\rm{A}}^{\rm{l}}D_{\rm{A}}^{\rm{ls}}}{2D_{\rm{A}}^{\rm{s}}}\left(\frac{\partial\alpha_x}{\partial x}+\frac{\partial\alpha_y}{\partial y}\right)\,.
\end{equation}

\subsection{Measuring Magnification in Galaxy Surveys}\label{sec:mu_dm}

In galaxy surveys, magnification can be quantified by observing variations in the number density or flux of the background sources. There are three effects that magnification induces on a background galaxy sample. Firstly, the flux of galaxies is magnified by $\mu$. Secondly, the number density of galaxies decreases by a factor of $\mu$, as the area is amplified by $\mu$. Finally, the intrinsic magnitude limit of the galaxy sample shifts towards fainter values, resulting in more galaxies being brought above the magnitude limit.

In prior research, changes in the number density $\delta n/n$ or the mean magnitude $\delta m$ of the background sources have typically been employed as observables for magnification. The relationships between $\delta n/n$, $\delta m$, and $\delta \mu$ in the weak lensing regime ($\delta \mu\ll1$) have been extensively formulated in the literature for magnitude-limited samples  \citep{2005ApJ...633..589S,2010MNRAS.405.1025M,2014MNRAS.440.3701B}:

\begin{align}
    \frac{\delta n}{n}&\approx \alpha_c\delta\mu\,,\notag\\ \alpha_c&=2.5\frac{d\log_{10}n(<m)}{dm}(m_{\rm{lim}})-1.0\,,
\end{align}
and
\begin{align}
    \delta m&\approx\alpha_m \delta\mu\,,\notag\\
    \alpha_m&=\frac{2.5}{\ln(10)}\left[-1+\frac{d\bar{m}(<m)}{dm}(m_{\rm{lim}})\right]\,,
\end{align}
where $m_{\rm{lim}}$ represents the magnitude limit of the galaxy sample, $n(<m)$ is the total number density of galaxies brighter than $m$, and $\bar{m}(<m)$ is the mean brightness of galaxies brighter than $m$.

In this study, however, we opt to measure the change in the total flux density $F$ of source galaxies around the lens to explore magnification. This choice avoids the need to measure the number density of galaxies required for computing $\delta n/n$ and $\delta m$. In crowded regions, the extraction of sources in the images may be impacted by the imperfect deblending problem. \citet{2021ApJ...919...25W} discovered that deblending mistakes can significantly influence the count of galaxies around central bright galaxies at scales where $r_{\rm{p}}<0.1R_{200}$\footnote{$R_{200}$ is defined as the radius within which the average matter density is 200 times the mean critical density of the universe.} in the Hyper Suprime-Cam (HSC) survey \citep{2018PASJ...70S...4A}. Although our source sample is from a survey shallower than HSC, this issue remains a concern. Hence, we opt to measure the total flux density $F$, which may be the most robust against the imperfect deblending issue. For instance, in crowded regions where a source might fragment into several components due to a higher image background, the number density could change substantially while the total flux density $F$ remains constant. In practice, we indeed find that the magnification results from $F$ are more accurate than those from $\delta n/n$ and $\delta m$.

The total flux density of a magnitude-limited sample can be expressed as:
\begin{equation}
    F(m_{\rm{lim}})=\int_{f_{\rm{lim}}}^{\infty}dffn(f)\,,
\end{equation}
where $f_{\rm{lim}}=10^{-m_{\rm{lim}}/2.5}$. When magnified by a lens, the first and second effects cancel out, as the flux of each source increases by $\mu$, while the number density decreases by $\mu$. The change in $F$ is solely induced by alterations in the magnitude limit of the sample:
\begin{align}
    \frac{F_{\mu}}{F}&=\frac{F(m_{\rm{lim}}+2.5\log_{10}\mu)}{F(m_{\rm{lim}})}\,.
\end{align}
Expressing this in the convention of magnitude, we have
\begin{align}
    \delta M &= -2.5\log_{10}\left(\frac{F_{\mu}}{F}\right)\notag\\
    &= -2.5\log_{10}{F_{\mu}}+2.5\log_{10}{F}\notag\\
    &= M(m_{\rm{lim}}+2.5\log_{10}\mu)-M(m_{\rm{lim}})\,.\label{eq:dm}
\end{align}
Where $M(m)$ is the total magnitude of galaxies brighter than $m$. In the weak lensing region where $\delta \mu\ll1$, $\delta M$ can be approximated by
\begin{align}
    \delta M &\approx \alpha_M\delta \mu\,,\notag\\
    \alpha_M &= \frac{2.5}{\ln(10)}\frac{dM}{dm}(m_{\rm{lim}})\,.\label{eq:lin_re}
\end{align}

In regions where $\delta \mu$ becomes significant, Equation \ref{eq:lin_re} becomes inaccurate for describing the $\delta\mu-\delta M$ relation. Fortunately, modern cosmological photometric surveys typically conduct small-area deep field surveys, which are much deeper than the wide surveys used for lensing studies. The $\delta\mu-\delta M$ relation can be empirically derived from a deeper sample by selecting various $\mu$ values and computing $M(m_{\rm{lim}}+2.5\log_{10}\mu)$ for the respective magnitude limits.

Moreover, in a more complex scenario where the source sample is already incomplete at the magnitude limit chosen for measuring magnification, the $\delta\mu-\delta M$ relation can still be derived from a deeper sample. For instance, in this study, the completeness of our source samples ranges from $90\%$ to $95\%$ at the chosen fiducial magnitude limits. In this case, $M(m_{\rm{lim}})$ in the source sample can be different from that in the deeper sample. Hence, we account for a more realistic $\delta\mu-\delta M$ relation by considering this incompleteness. The completeness, represented as a function of flux $C(f)$ or magnitude $C(m)$, can be determined by comparing the galaxy number density in the source sample to that in the deeper sample. Subsequently, the total flux density can be re-write as
\begin{equation}
    F(m_{\rm{lim}})=\int_{f_{\rm{lim}}}^{\infty}dffC(f)n(f)\,.
\end{equation}
Assuming that magnification does not affect the incompleteness for a fixed magnitude, as the survey depth remains unaffected by magnification, the magnified total flux density can be obtained from the deeper sample:
\begin{equation}
    F_{\mu}(m_{\rm{lim}},\mu)=\int_{f_{\rm{lim}}/\mu}^{\infty}dffC(\mu f)n(f)\,.\label{eq:incom}
\end{equation}
Then, $\delta M$ for a specific $\mu$ can be obtained by comparing $M(m_{\rm{lim}})$ obtained from the source sample with $M_{\mu}(m_{\rm{lim}},\mu)$ derived from the deeper sample. It is also worth noting that if $C(f)$ varies quite slowly across the magnitude ranges we are interested in, such that $dC/df\approx0$, the $\delta\mu-\delta M$ relation approaches one that does not account for incompleteness.

\section{Observational data}\label{sec:obs_data}

In this section, we describe the lens and source samples employed in our observational measurements.

\begin{figure}
    \plotone{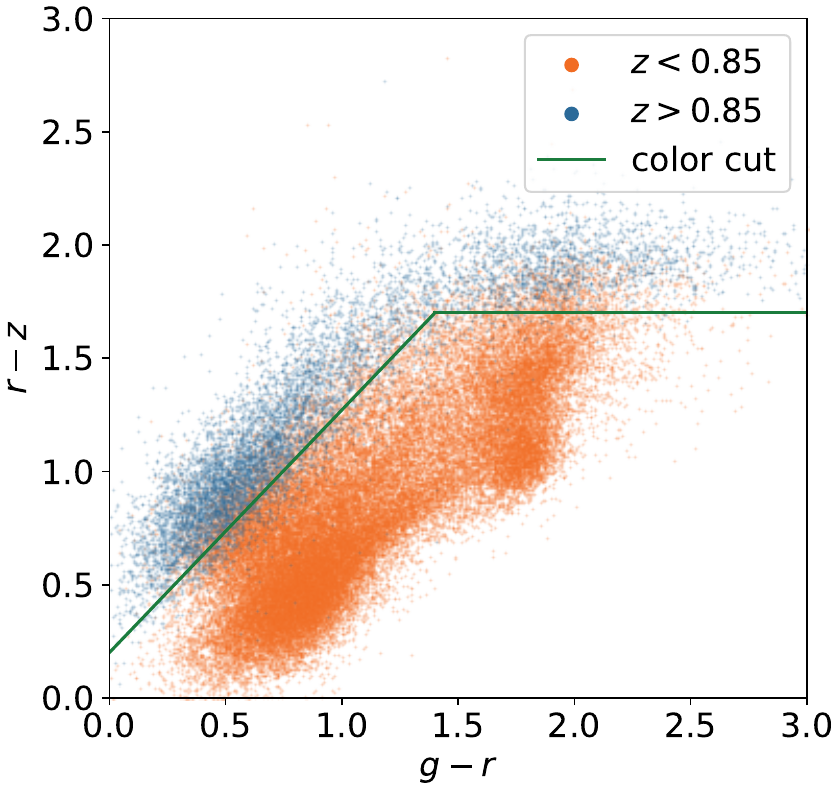}
    \caption{The $g-r$ vs. $r-z$ diagram for the VIPERS sample. Blue dots represent galaxies with $z>0.85$, while orange dots indicate those with $z<0.85$. Green lines delineate the color cut employed to select the background source sample with $z>0.85$.}
    \label{fig:color_cut}
\end{figure}

\subsection{Lens Sample}\label{sec:lens}
For the lens sample, we utilize the CMASS spectroscopic sample from SDSS-III BOSS DR12 \citep{2015ApJS..219...12A,2016MNRAS.455.1553R}. The CMASS sample employs target selections similar to those of the SDSS-I/II Cut II Large Red Galaxy (LRG) sample but is bluer and fainter to increase the galaxy number density in the redshift range of $0.43<z<0.75$. Galaxies in the CMASS sample are selected using a combination of magnitude and color cuts to achieve an approximately constant stellar mass. The magnitude limits for the {\tt{cmodel}} magnitudes in the CMASS sample are $17.6<i<19.9$, and the full selection criteria can be found in \citet{2016MNRAS.455.1553R}. We utilize the "{\tt{CMASS}}" LSS catalog\footnote{https://data.sdss.org/sas/dr12/boss/lss/} from BOSS DR12 for the CMASS sample, which covers $9376\ \rm{deg}^2$ of the sky, with $6851\ \rm{deg}^2$ in the NGC and $2525\ \rm{deg}^2$ in the SGC. To match the footprint of the source sample described in Section \ref{sec:source}, we adopt an angular cut of $\rm{decl.}\leq32^{\circ}$, resulting in an effective area of approximately $6600\ \rm{deg}^2$. To ensure a sufficient redshift gap between the lens and source samples while including as many lens galaxies as possible, we apply a redshift cut of $0.5<z<0.65$. This results in a final sample of 348,938 LRGs in our lens sample.

\begin{figure}
    \plotone{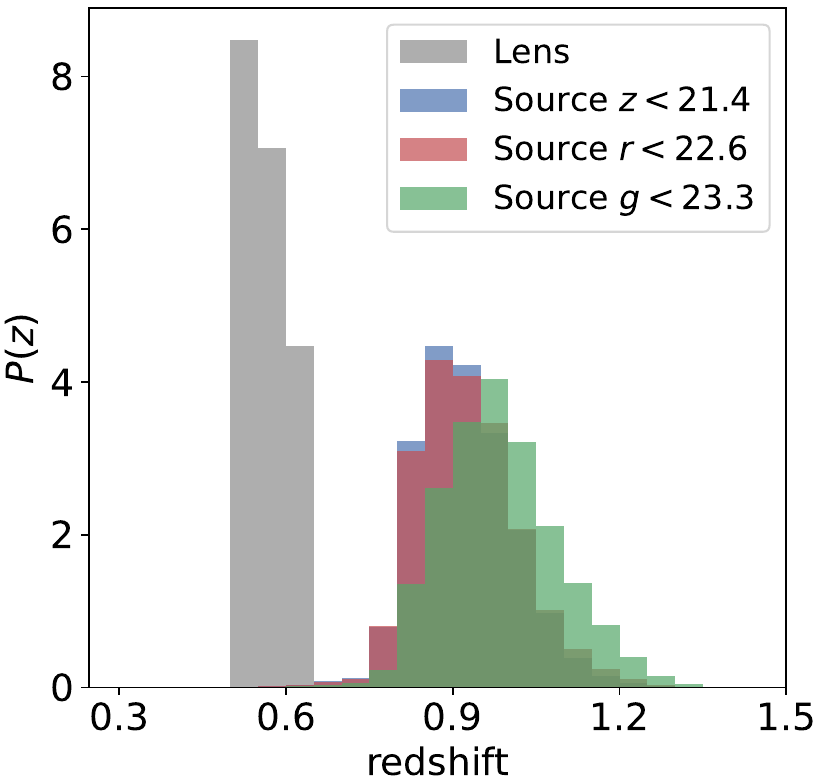}
    \caption{The redshift distributions of the CMASS lens sample and three DECaLS source samples with different magnitude limits in the $grz$ bands are presented. The CMASS sample is depicted with spectroscopic redshifts, while DECaLS samples are represented using photoz.}
    \label{fig:redshift}
\end{figure}

\subsection{Source Sample}\label{sec:source}
For the source sample, we utilize the DECaLS photometric catalog\footnote{https://www.legacysurvey.org/dr9/catalogs/} from the DR9 of the DESI Legacy Imaging Surveys \citep[Legacy Surveys; ][]{2019AJ....157..168D}. Covering approximately $9000\ {\rm{deg}}^2$ in both the Northern and Southern Galactic caps (NGC and SGC) at $\rm{decl.}\leq32^{\circ}$, DECaLS provides imaging data in $g$, $r$, and $z$ bands, with median $5\sigma$ point source depths of $24.9$, $24.2$, and $23.3$ respectively. Additionally, DECaLS incorporates data from the deeper Dark Energy Survey (DES; \citealt{2016MNRAS.460.1270D}), extending coverage by an additional $5000\ {\rm{deg}}^2$ in the SGC. Image processing is conducted using {\texttt{Tractor}} \citep{2016ascl.soft04008L} for source extraction, employing parametric profiles convolved with specific point spread functions (PSFs), including a delta function for point sources, exponential and de Vaucouleurs laws, as well as S\'ersic profiles. The images from DES have been reprocessed using the same pipeline to ensure consistency. Throughout the analysis, we utilize the best-fit model magnitudes provided by {\texttt{Tractor}}. We restrict our analysis to footprints observed at least once in all three bands and apply a bright star mask and bad pixel mask using the MASKBITS\footnote{https://www.legacysurvey.org/dr9/bitmasks/} provided by the Legacy Surveys. Additional masks are employed to match the geometry of the lens sample described in Section \ref{sec:lens}. The final footprint covers approximately $6600\ \rm{deg}^2$. Galactic extinction correction is applied to all sources using the maps of \citet{1998ApJ...500..525S}. To exclude stars, sources with point source (PSF) morphologies are removed from the sample.

\begin{figure*}
\centering
    \includegraphics[width=\textwidth]{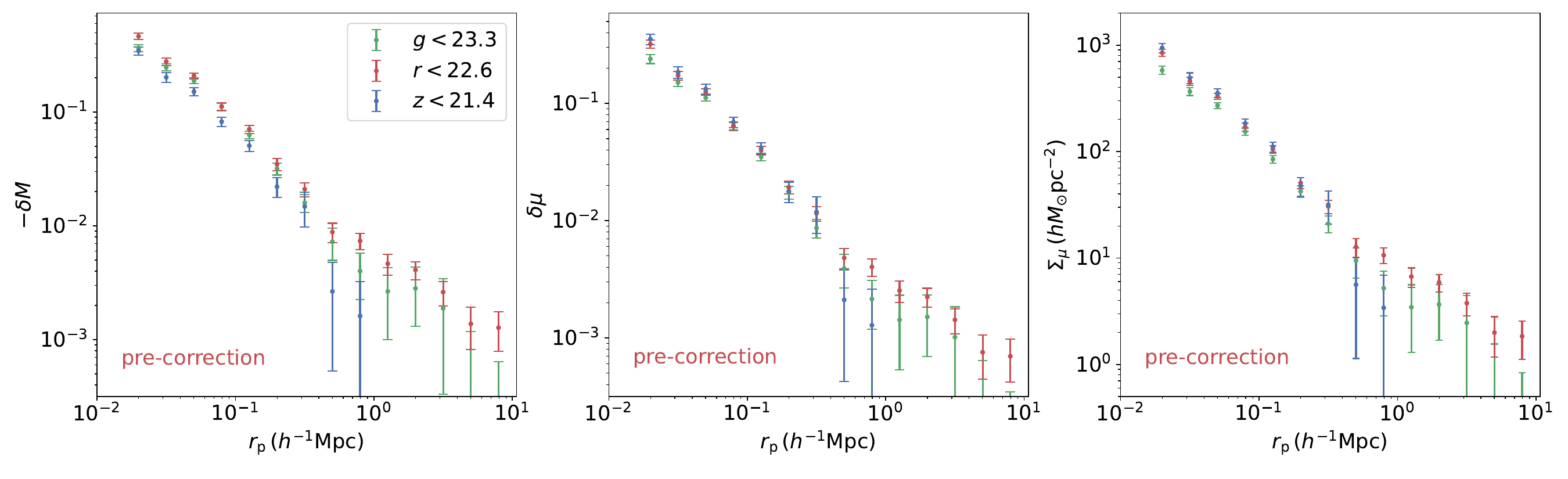}
    \caption{Radial distributions of $\delta M$, $\delta \mu$, and $\Sigma_{\mu}$ measurements around lens galaxies across three bands before the corrections for dust attenuation.}
    \label{fig:measure_pre}
\end{figure*}

To construct a background source sample, our goal is to select sources with redshifts higher than $0.85$ to ensure an adequate redshift separation from the lens sample. Rather than relying on photometric redshifts (photoz), we employ a simple color cut to avoid potential complications associated with unknown and complex selection criteria inherent in photoz. We employ the spectroscopic galaxy catalog from the final public release (PDR-2) of the VIMOS Public Extragalactic Redshift Survey \citep[VIPERS; ][]{2018A&A...609A..84S} to define the color cut. VIPERS is a magnitude-limited spectroscopic survey spanning approximately 24 ${\rm{deg}}^2$, targeting galaxies with $i<22.5$ and incorporating an extra color criterion to eliminate low-redshift galaxies ($z<0.5$). We match the VIPERS sample with DECaLS to get the same $grz$ band flux measurements. In Figure \ref{fig:color_cut}, we plot the VIPERS galaxies with $z<0.85$ and $z>0.85$ in the $g-r$ vs. $r-z$ diagram and find that the following color cut can effectively reject galaxies with $z<0.85$:
\begin{equation}
r-z>
\begin{cases}
    \frac{15}{14}(g-r)+0.2 & (g-r<1.4)\,,\\
    1.7 & (g-r\geq1.4)\,.\label{eq:color_cut}
\end{cases}
\end{equation}

After extracting high-redshift sources using the color cut described in Equation \ref{eq:color_cut}, we select three source samples from DECaLS based on magnitude cuts in the $grz$ bands. These samples are utilized to measure dust attenuation and correct it for magnification results, as outlined in Section \ref{sec:dust_theory}. Our fiducial choose of the magnitude cuts are $g<23.3$, $r<22.6$ and $z<21.4$ according to the survey depths of DECaLS. We compare the fiducial results with those obtained using different magnitude cuts in Section \ref{sec:dust_corrected} and verify that our results remain consistent regardless of the choice of magnitude cuts. In Figure \ref{fig:redshift}, we present the redshift distribution of the CMASS lens sample and three DECaLS source samples. The CMASS sample is represented with spectroscopic redshifts, whereas the DECaLS samples are depicted using photoz from \citet{2021MNRAS.501.3309Z}. The lens sample has a mean redshift of $\bar{z}_{\rm{l}}=0.56$, while the source samples have mean redshifts of $\bar{z}_{\rm{s}}^g=0.98$, $\bar{z}_{\rm{s}}^r=0.93$, and $\bar{z}_{\rm{s}}^z=0.92$. The redshift distributions demonstrate the success of the color cut developed in Equation \ref{eq:color_cut} in selecting background samples.

\begin{figure*}
\centering
    \includegraphics[width=\textwidth]{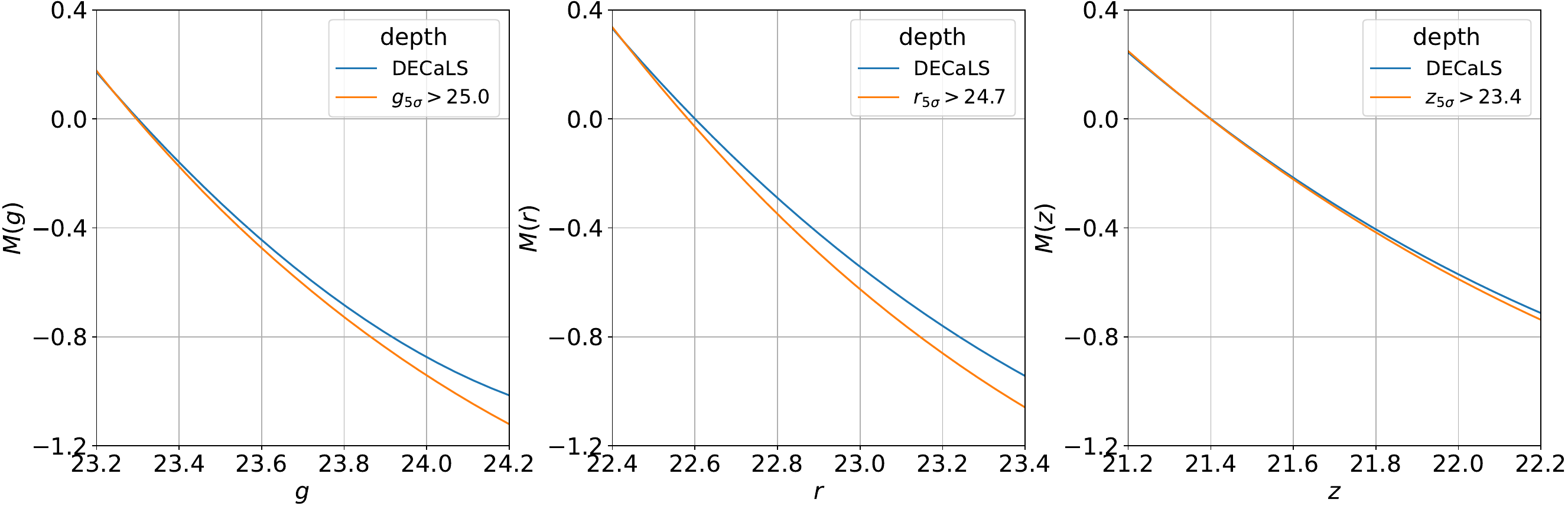}
    \caption{The distributions of the total magnitude $M$ in the $grz$ bands for both the DECaLS sample and the deeper samples. The zero points of $M$ are chosen based on the values from DECaLS at our fiducial magnitude cut of $g=23.3$, $r=22.6$, and $z=21.4$.}
    \label{fig:M_dis}
\end{figure*}

\begin{figure*}
\centering
    \includegraphics[width=\textwidth]{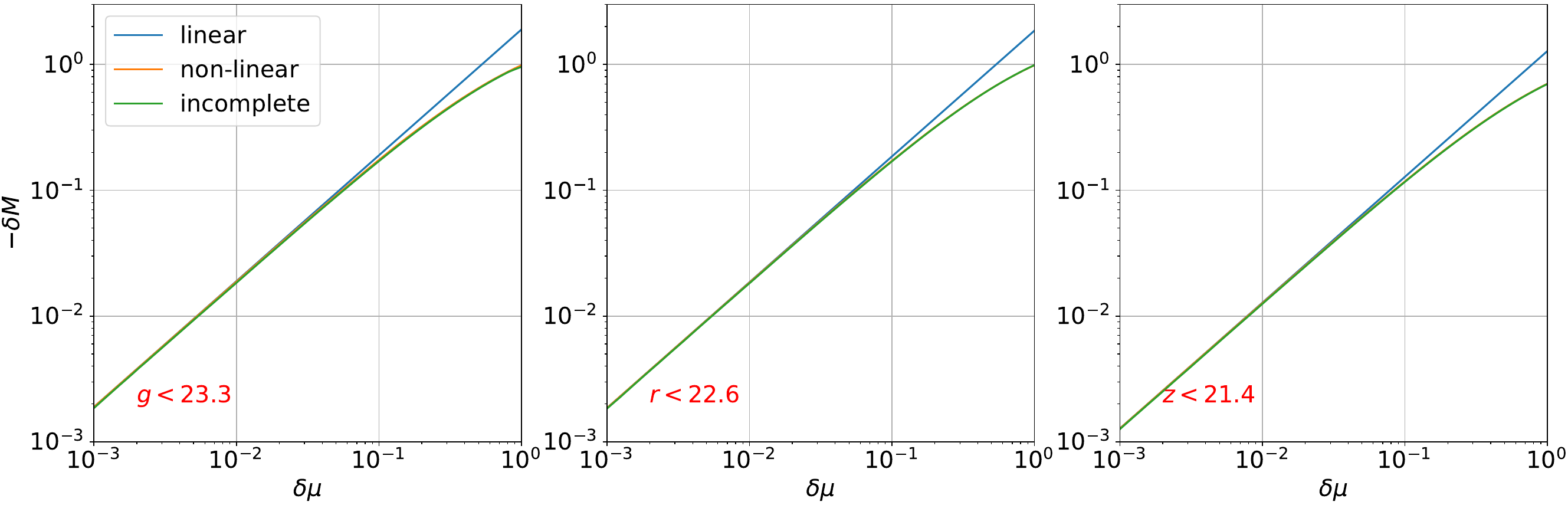}
    \caption{The $\delta \mu-\delta M$ relations $G_{\lambda}(\delta \mu)$ for $g<23.3$, $r<22.6$, and $z<21.4$. Blue lines depict the results from linear approximation in Equation \ref{eq:lin_re}. Orange lines represent the non-linear relations calculated directly from Equation \ref{eq:dm}. Green lines further consider the incompleteness of the samples as shown in Equation \ref{eq:incom}.}
    \label{fig:dmu_dm}
\end{figure*}

\section{Measuring magnification around CMASS galaxies}\label{sec:measurements}

In this section, we measure the magnification signal around the CMASS lens sample for the three fiducial source samples in different bands. We adopt the Planck18 cosmology \citep{2020A&A...641A...6P} with $\Omega_{\mathrm{m}} = 0.3111$, $\Omega_{\Lambda} = 0.6889$ and $H_0 = 67.66 \,\mathrm{km\,s^{-1}\,Mpc^{-1}}$ in the measurements.

\subsection{Measuring $\delta M$}
To begin, we estimate the radial distribution of the flux excess $F_{\mu}/F$ of the sources around lenses using a generalized Landy–Szalay estimator \citep{1993ApJ...412...64L}:
\begin{equation}
    \frac{F_{\mu}}{F}(r_{\rm{p}})=\frac{D_1F_2(r_{\rm{p}})-\bar{f}D_1R_2(r_{\rm{p}})-R_1F_2(r_{\rm{p}})}{\bar{f}R_1R_2(r_{\rm{p}})}+2\,,
\end{equation}
where $\bar{f}$ represents the mean flux of the source sample, $D_1$ and 
$R_1$ denote the number counts of the lens sample and its random sample respectively, and $F_2$ and $R_2$ stand for the fluxes of the source sample and the number counts of its random sample. For each lens galaxy, we employ $\theta=\arcsin{(r_{\rm{p}}/D_{\rm{A}}^l})$ to summarize the flux for the corresponding $r_{\rm{p}}$, where $D_{\rm{A}}^l$ represents the angular diameter distance to the lens galaxy and $r_{\rm{p}}$ is represented in the {\textit{physical}} coordinate system.

Then, the distribution of $\delta M$ can be calculated from $F_{\mu}/F$ by
\begin{equation}
    \delta M(r_{\rm{p}})=-2.5\log_{10}\left[\frac{F_{\mu}}{F}(r_{\rm{p}})\right]\,.
\end{equation}

For estimating the covariance matrix for $\delta M$, we employ the jackknife resampling method. The footprint of both NGC and SGC is partitioned into $N_{\rm{JK}}$ jackknife sub-samples. Subsequently, $\delta M$ is measured within each sub-sample. The final results of the $i$th sub-sample are obtained by combining measurements from the $i$th NGC sub-sample and the $i$th SGC sub-sample, weighted by the respective areas of NGC and SGC. The mean value and covariance matrix of $\delta M$ are then derived by
\begin{equation}
\delta\bar{M}(r_{\rm{p}})=\frac{1}{N_{\rm{JK}}}\sum_{i=1}^{N_{\rm{JK}}}\delta M_i(r_{\rm{p}})\,,
\end{equation}
\begin{align}
    C_{ab}^{M} = \frac{N_{\rm{JK}}-1}{N_{\rm{JK}}}\sum_{i=1}^{N_{\rm{JK}}}(\delta M_i(r_{\rm{p}}^a)-\delta\bar{M}(r_{\rm{p}}^a))\notag\\
    (\delta M_i(r_{\rm{p}}^b)-\delta\bar{M}(r_{\rm{p}}^b))\,.
\end{align}
where $a$ and $b$ denotes the $a$th and $b$th radial bins.

We measure the distributions of $\delta M$ around CMASS lens galaxies across three bands within the physical scales of $0.016\,h^{-1}{\rm{Mpc}}<r_{\rm{p}}<10\,h^{-1}{\rm{Mpc}}$, utilizing three magnitude-limited source samples. We set $N_{\rm{JK}}=200$ in our study. The results are depicted in the left panel of Figure \ref{fig:measure_pre}. We observe an increase in the signal-to-noise ratio (S/N) of $\delta M$ with decreasing $r_{\rm{p}}$, achieving highly accurate measurements down to $20\,h^{-1}{\rm{kpc}}$, underscoring the potential of magnification for exploring matter distribution within the inner regions of dark matter halos.

\subsection{Establishing the $\delta\mu-\delta M$ Relations}
After obtaining measurements of $\delta M$, establishing the $\delta\mu-\delta M$ relation becomes necessary to convert the observed quantities into physical quantities, facilitating easier comparisons with models. As delineated in Section \ref{sec:mu_dm}, obtaining a deeper sample is crucial for deriving the $\delta\mu-\delta M$ relation, particularly as we delve into the non-linear region. Fortunately, DECaLS incorporates data from DES, which extends survey depths by at least 0.5 magnitude across all three bands. Consequently, we choose the deepest regions within the DES area to investigate the $\delta\mu-\delta M$ relations, with $5\sigma$ PSF depths exceeding 25.0, 24.7, and 23.4 for the $g$, $r$, and $z$ bands, respectively. The areas covered by these deeper samples are 371 deg$^2$, 649 deg$^2$, and 1249 deg$^2$, respectively, which are large enough to obtain accurate measurements of the total flux density.

As demonstrated by Equation \ref{eq:dm}, the $\delta\mu-\delta M$ relation is determined by the distribution of the total magnitude $M$. In Figure \ref{fig:M_dis}, we depict the distributions of $M$ for the three bands, encompassing both the DECaLS sample and deeper samples. For better representation, we align the zero points of $M$ with the values of the DECaLS sample at the fiducial magnitude cuts, as only the relative changes of $M$ matter. It is clear that the values of $M$ for the deeper samples decrease in comparison to DECaLS as we move towards fainter magnitudes. This suggests that the DECaLS sample becomes increasingly incomplete for fainter magnitude limits. From Figure \ref{fig:M_dis}, we observe that the DECaLS sample is already slightly incomplete at the fiducial magnitude limits. This could potentially affect the $\delta\mu-\delta M$ relation, as explored in Section \ref{sec:mu_dm}, and should be duly considered.

Using the distributions of $M$ for DECaLS and deeper samples, we establish the $\delta\mu-\delta M$ relations for $g<23.3$, $r<22.6$, and $z<21.4$. In Figure \ref{fig:dmu_dm}, we present the $\delta\mu-\delta M$ relations for the three magnitude-limited samples. For each sample, we derive three distinct $\delta\mu-\delta M$ relations: a linear relation (Equation \ref{eq:lin_re}), a non-linear relation (Equation \ref{eq:dm}), and a non-linear relation accounting for incompleteness (Equation \ref{eq:incom}). We observe that non-linear effects become prominent when $\delta \mu$ reaches $0.1$. Furthermore, we find minimal alteration in the relations after accounting for incompleteness, suggesting that the slight incompleteness in our fiducial samples has negligible impact on the $\delta\mu-\delta M$ relations.

\begin{figure}
    \plotone{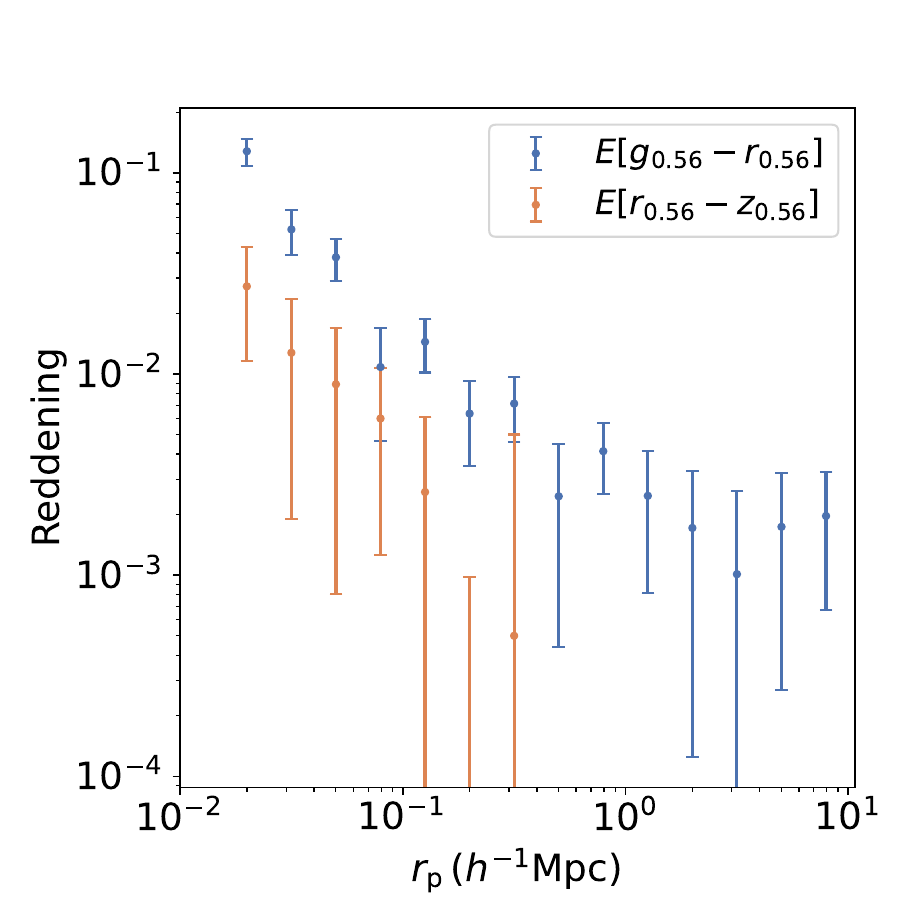}
    \caption{The distributions of $E[g_{0.56}-r_{0.56}]$ and $E[r_{0.56}-z_{0.56}]$  around CMASS lens galaxies.}
    \label{fig:reddening}
\end{figure}

\begin{figure*}
\centering
    \includegraphics[width=\textwidth]{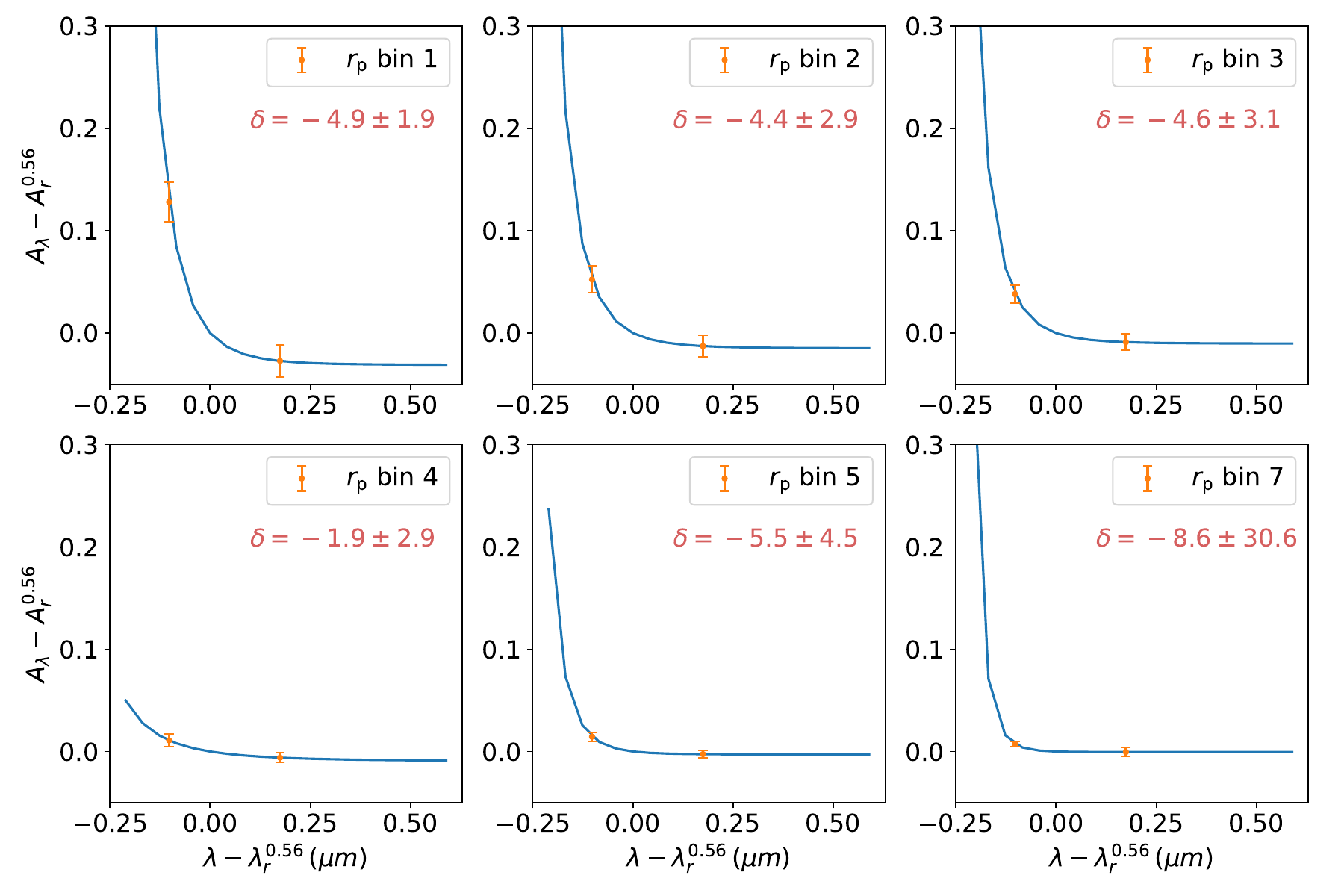}
    \caption{The fits of $E[g_{0.56}-r_{0.56}]$ and $E[r_{0.56}-z_{0.56}]$ using the modified \citet{2000ApJ...533..682C} attenuation curves in 6 $r_{\rm{p}}$ bins.}
    \label{fig:fit_dust_two}
\end{figure*}

\begin{figure*}
\centering
    \includegraphics[width=\textwidth]{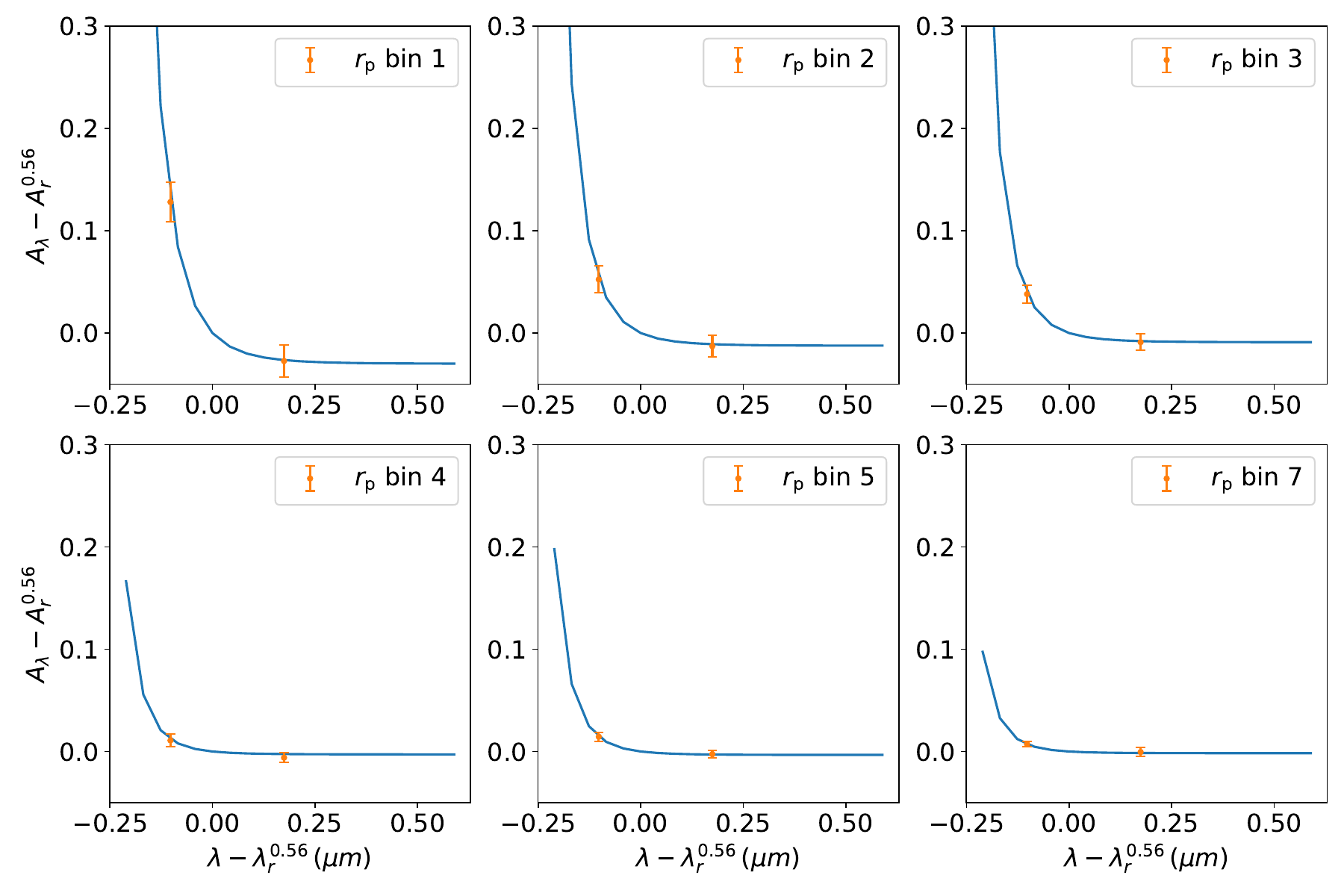}
    \caption{Similar to Figure \ref{fig:fit_dust_two} but with $\delta$ set to $-5$.}
    \label{fig:fit_dust_one}
\end{figure*}
\begin{figure*}
\centering
    \includegraphics[width=\textwidth]{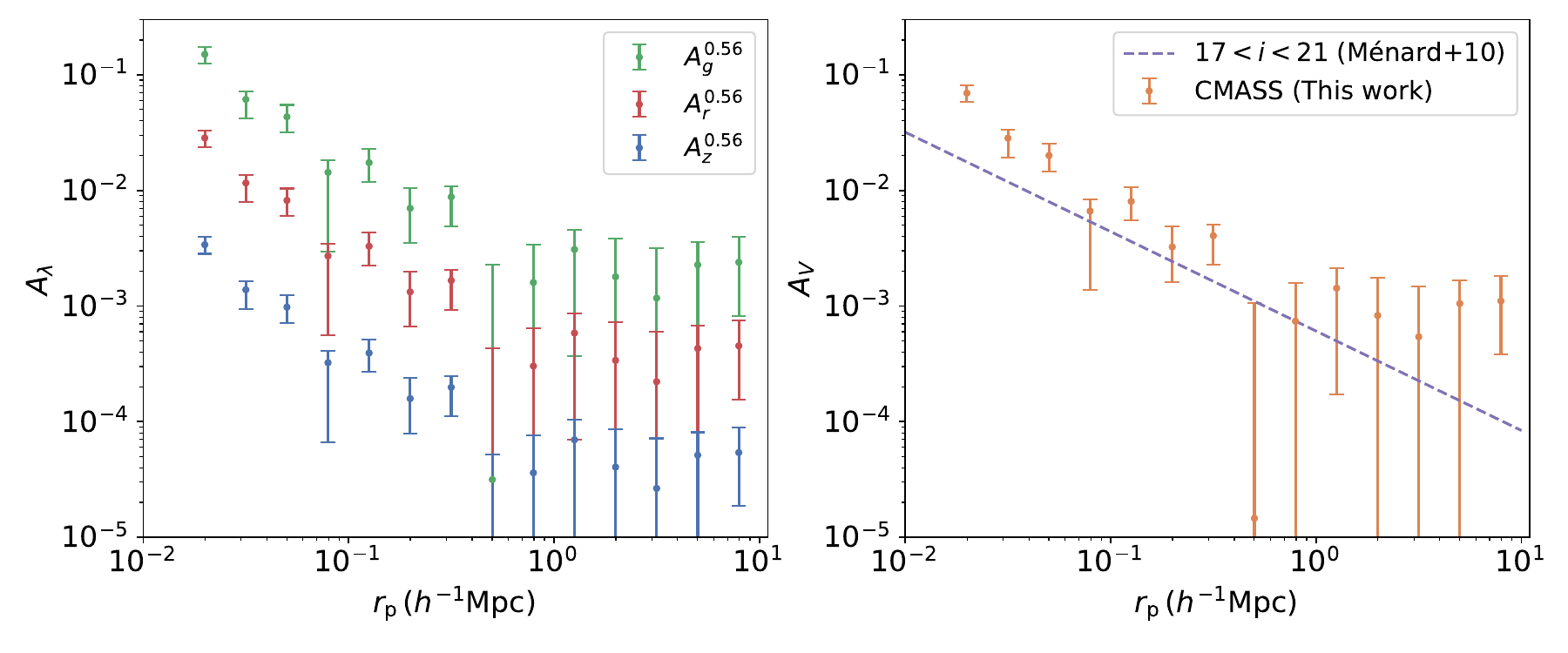}
    \caption{The dust attenuation in various bands around CMASS lens galaxies. Left: The dust attenuation in the $grz$ bands at a redshift around 0.56. Right: The $V$ band attenuation around CMASS galaxies compared to the average attenuation in the $V$ band from \citet{2010MNRAS.405.1025M} around all galaxies with $17<i<21$.}
    \label{fig:A_lambda}
\end{figure*}

\subsection{Obtaining $\delta \mu$ and $\Sigma_{\mu}$}
Using the $\delta\mu-\delta M$ relationships, we can translate the measured $\delta M$ into the more physical quantities $\delta \mu$ and $\Sigma_\mu$. The non-linear relations accounting for incompleteness are used in this study. We employ a similar approach as for $\delta M$ to determine the mean value and covariance matrix of $\delta\mu$. Initially, we convert $\delta M_i$ in each jackknife sub-sample into $\delta\mu_i$, and then calculate:
\begin{equation}
\delta\bar{\mu}(r_{\rm{p}})=\frac{1}{N_{\rm{JK}}}\sum_{i=1}^{N_{\rm{JK}}}\delta \mu_i(r_{\rm{p}})\,,
\end{equation}
\begin{align}
    C_{ab}^{\mu} = \frac{N_{\rm{JK}}-1}{N_{\rm{JK}}}\sum_{i=1}^{N_{\rm{JK}}}(\delta \mu_i(r_{\rm{p}}^a)-\delta\bar{\mu}(r_{\rm{p}}^a))\notag\\
    (\delta \mu_i(r_{\rm{p}}^b)-\delta\bar{\mu}(r_{\rm{p}}^b))\,.
\end{align}
Subsequently, we determine the mean value and covariance matrix of $\Sigma_\mu$ by dividing those of $\delta\mu$ by $\Sigma_{\rm{crit}}/2$, where $\Sigma_{\rm{crit}}$ is calculated using the mean redshifts of the lens and source samples.

The distributions of $\delta \mu$ and $\Sigma_{\mu}$ around the CMASS lens sample for our fiducial source samples in three bands are depicted in the middle and right panels of Figure \ref{fig:measure_pre}. Ideally, $\Sigma_{\mu}$ should remain consistent across the three bands. However, we observe that longer-wavelength bands exhibit higher $\Sigma_{\mu}$, suggesting the potential presence of dust in the CGM of the massive CMASS galaxies. Dust attenuation can be corrected by comparing the measured $\Sigma_{\mu}$ in three bands, which will be investigated in next section.

\section{Dust attenuation around CMASS galaxies}\label{sec:dust}
In this section, we compare the \(\Sigma_{\mu,\lambda}^{\rm{obs}}\) measurements in the \(grz\) bands to derive the reddening effects around CMASS lens galaxies and infer the dust attenuation curves. We denote the observed \(\Sigma_{\mu}\) before dust correction in different bands as \(\Sigma_{\mu,\lambda}^{\rm{obs}}\). Subsequently, we correct for dust attenuation in our measurements to obtain the true magnification signals.

\subsection{Dust Attenuation in the Measurements of Magnification }\label{sec:dust_theory}
When light from background sources passes through lens galaxies, it can be not only magnified but also absorbed by the dust within the lenses \citep{2010MNRAS.405.1025M}. These two effects become degenerate when measuring $\delta M$:
\begin{equation}
    \delta M_{\lambda}^{\rm{obs}} = \delta M_{\lambda}+A_{\lambda}\,,
\end{equation}
where $A_{\lambda}$ denotes the magnitude change at wavelength $\lambda$ due to dust attenuation. 

Fortunately, $A_{\lambda}$ changes with wavelength, as described by the dust attenuation curve \citep{2000ApJ...533..682C,2000ApJ...539..718C}, while $\mu$ remains constant for the same sample across all wavelengths. Leveraging this distinction, dust attenuation can be measured by comparing magnification results across different observed bands, allowing extraction of the true magnification signal after correcting for dust attenuation.

However, measuring dust attenuation from the same sample may be challenging since only one band is magnitude limited, while others can be subject to complicated selection criteria, making it difficult to establish the $\delta\mu-\delta M$ relation for those bands. Instead, in this study, as show in Section \ref{sec:source}, we select a magnitude-limited source sample in each band to measure dust attenuation. In this scenario, $\mu$ begins to vary for different samples as they may have different redshift distributions. In this case, the quantity that remains unchanged is the matter surface density $\Sigma$ of the lenses. 

Let us denote the $\delta\mu-\delta M$ relation shown in Figure \ref{fig:dmu_dm} as $\delta M=G_{\lambda}(\delta \mu)$ for better representation. In the weak lensing regime ($\delta \mu\ll1$), the observed magnitude change of sources around lenses for the sample in each band is given by:
\begin{equation}
    \delta M_{\lambda}^{\rm{obs}}(\bm{\theta}) = G_{\lambda}\left(\frac{2\Sigma(\bm{\theta})}{\Sigma_{\rm{crit},\lambda}}\right)+A_{\lambda}(\bm{\theta})\,.\label{eq:mag_p_dust_lin}
\end{equation}
With a parameterized dust attenuation law, we can fit $\delta M_{\lambda}^{\rm{obs}}$ in different bands and constrain the matter surface density and dust attenuation simultaneously. 

In the non-linear regime, we replace $\Sigma$ in Equation \ref{eq:mag_p_dust_lin} by $\Sigma_{\mu}$:
\begin{equation}
    \delta M_{\lambda}^{\rm{obs}}(\bm{\theta}) = G_{\lambda}\left(\frac{2\Sigma_{\mu}(\bm{\theta})}{\Sigma_{\rm{crit},\lambda}}\right)+A_{\lambda}(\bm{\theta})\,.\label{eq:mag_p_dust}
\end{equation}
As mentioned in Section \ref{sec:basics}, \(\Sigma_{\mu}\) is theoretically different for samples selected in different bands since it still weakly depends on the source redshift distribution. However, for our fiducial source samples with very close mean redshifts, we find that the remaining redshift dependence is negligible ($<3\%$), as demonstrated in Section \ref{sec:simu_lens}.

\subsection{Reddening}
In the ideal scenario where we possess complete knowledge of the dust attenuation curve, it would be straightforward to fit the measurements directly using Equation \ref{eq:mag_p_dust} to simultaneously constrain the true $\Sigma_{\mu}$ and dust attenuation curve $A_{\lambda}$. However, the dust attenuation curve can vary significantly in different environments and lacks a universal parameterization. Therefore, for a more realistic approach, we first measure the reddening $E[g_{0.56}-r_{0.56}]$ and $E[r_{0.56}-z_{0.56}]$ to obtain the {\it approximate shape} of the dust attenuation curve, where the subscription 0.56 indicates the mean redshifts of the lenses.

To obtain \(E[g_{0.56}-r_{0.56}]\), we compute the values of \(\delta M_g\) required for the \(g\)-band to achieve the \(\Sigma_{\mu,r}^{\rm{obs}}\) measured in the \(r\)-band and compare them with the observed values \(\delta M_{g}^{\rm{obs}}\):
\begin{align}
 \label{eq:red1}
    E[g_{0.56}-r_{0.56}](\theta) &= A_g(\theta) - A_r(\theta) \notag \\
    &\approx G_g\left(\frac{2\Sigma_{\mu,r}^{\rm{obs}}}{\Sigma_{{\rm{crit}},g}}\right) - \delta M_{g}^{\rm{obs}}\,.
\end{align}
Similarly, we can obtain \(E[r_{0.56}-z_{0.56}]\):
\begin{equation}
    E[r_{0.56}-z_{0.56}](\theta) \approx G_r\left(\frac{2\Sigma_{\mu,z}^{\rm{obs}}}{\Sigma_{{\rm{crit}},r}}\right) - \delta M_{r}^{\rm{obs}}\,.
\end{equation}
Here, \(\Sigma_{{\rm{crit}},g}\) and \(\Sigma_{{\rm{crit}},r}\) are the \(\Sigma_{{\rm{crit}}}\) for \(g\) and \(r\) band source samples, and $G_g(\delta \mu)$ and $G_r(\delta \mu)$ are represented in Figure \ref{fig:dmu_dm}. Note that this method only provides a {\it leading-order} approximation of reddening rather than the exact true values. However, we believe this should be sufficient to investigate the form of the attenuation curve. After determining the form of the attenuation curve, we will derive the exact dust attenuation according to Equation \ref{eq:mag_p_dust}.

We compute $E[g_{0.56}-r_{0.56}]$ and $E[r_{0.56}-z_{0.56}]$ in each jackknife sub-sample and derive their mean values and covariance matrices. The distributions of $E[g_{0.56}-r_{0.56}]$ and $E[r_{0.56}-z_{0.56}]$ around the CMASS lens sample are shown in Figure \ref{fig:reddening}. We observe that $E[g_{0.56}-r_{0.56}]$ and $E[r_{0.56}-z_{0.56}]$ decrease with $r_{\rm{p}}$, suggesting a potential correlation between the dust distribution and the matter distribution. Additionally, we note that $E[g_{0.56}-r_{0.56}]$ is substantially larger than $E[r_{0.56}-z_{0.56}]$, indicating a rapid change in the slope of the dust attenuation curve, which can help constrain the shape of the dust attenuation curve.

\subsection{Dust Attenuation Curves}
Comparing our measurements to the commonly adopted starburst attenuation curve from \citet{2000ApJ...533..682C}, we observe a more rapid decrease from $E[g_{0.56}-r_{0.56}]$ to $E[r_{0.56}-z_{0.56}]$. Therefore, we adopt a more flexible form for the attenuation curve to analysis our measurements. We utilize the modified \citet{2000ApJ...533..682C} model \citep{2009A&A...507.1793N} to fit $E[g_{0.56}-r_{0.56}]$ and $E[r_{0.56}-z_{0.56}]$:
\begin{equation}
    A(\lambda)=k_{\rm{C00}}\left(\frac{\lambda}{550\,{\rm{nm}}}\right)^{\delta}\frac{E(B-V)_{\delta=0}}{E(B-V)_{\delta}},\label{eq:atte_curve}
\end{equation}
where $k_{\rm{C00}}$ represents the attenuation curve in \citet{2000ApJ...533..682C}, $550\, {\rm{nm}}$ denotes the effective wavelength of the $V$ band used as reference, and the last term serves for renormalization. The modified model comprises two free parameters, among which $\delta$ alters the shape of the attenuation curve, while all the others are degenerate and determine the amplitude. 

We fit the measured $E[g_{0.56}-r_{0.56}]$ to $E[r_{0.56}-z_{0.56}]$ for the first 5 $r_{\rm{p}}$ bins and the 7th $r_{\rm{p}}$ bin, as the $z$-band measurements for other bins are not good and we lack reasonable $E[r_{0.56}-z_{0.56}]$ measurements, as shown in Figure \ref{fig:reddening}. The effective wavelengths of the DECaLS $grz$ bands are 482 nm, 642 nm, and 915 nm, respectively. We shift them to redshift $0.56$ to derive the effective wavelengths for $g_{0.56}$, $r_{0.56}$, and $z_{0.56}$, which are then utilized in the fitting process. The fitting results are presented in Figure \ref{fig:fit_dust_two}. We observed that very steep attenuation curves with $\delta\approx-5$ are required to explain our measurements.

To correct the dust attenuation effects in our magnification measurements, we require the attenuation curves for all $r_{\rm{p}}$ bins. While measurements of $E[r_{0.56}-z_{0.56}]$ are lacking for other bins, we do possess reliable measurements of $E[g_{0.56}-r_{0.56}]$ across all scales. However, to constrain the attenuation curve, we must reduce one degree of freedom since the modified \citet{2000ApJ...533..682C} model contains two free parameters. Thus, based on the findings in Figure \ref{fig:fit_dust_two}, we set $\delta=-5$ and re-fit the 6 $r_{\rm{p}}$ bins to assess the validity of this choice. The fits with $\delta=-5$ are illustrated in Figure \ref{fig:fit_dust_two}. We observe that the modified \citet{2000ApJ...533..682C} model with $\delta=-5$ effectively fits the measurements across all 6 $r_{\rm{p}}$ bins. Therefore, it appears reasonable to extend the application of this form to other scales.

\begin{figure*}
\centering
    \includegraphics[width=\textwidth]{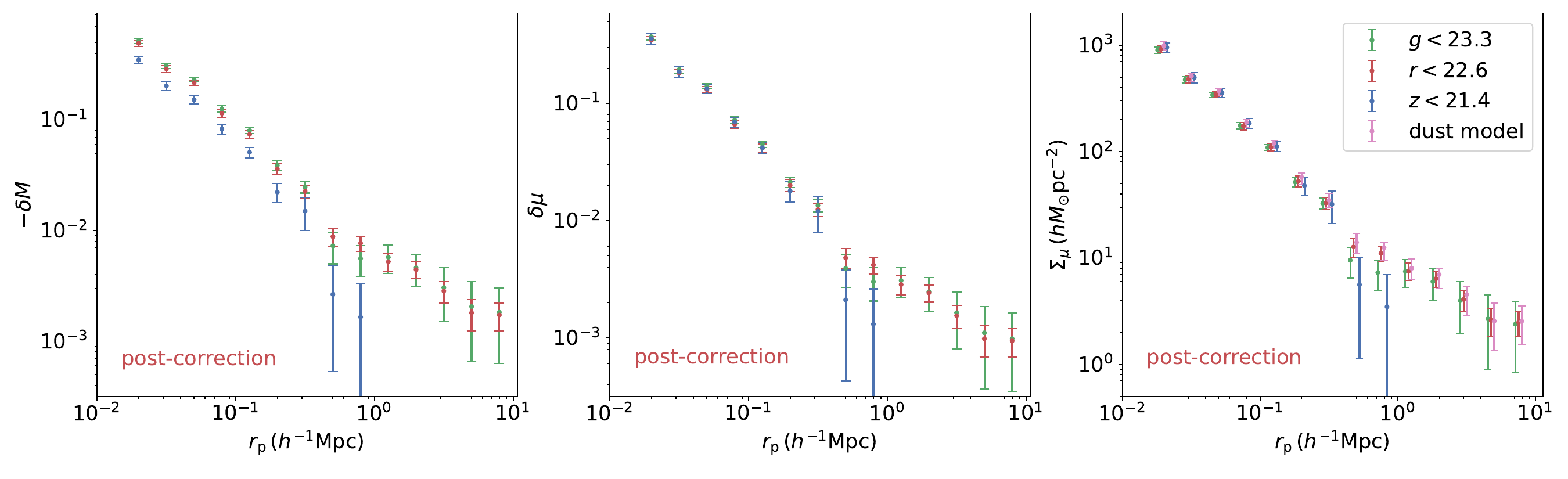}
    \caption{Similar to Figure \ref{fig:measure_pre}, but with corrections for dust attenuation. The combined constraints of $\Sigma_{\mu}$ from the dust model (Equation \ref{eq:dust_chi2}) are also depicted in the right panel. The results in the right panel are horizontally shifted for better illustration.}
    \label{fig:measure_post}
\end{figure*}

\begin{figure*}
\centering
    \includegraphics[width=\textwidth]{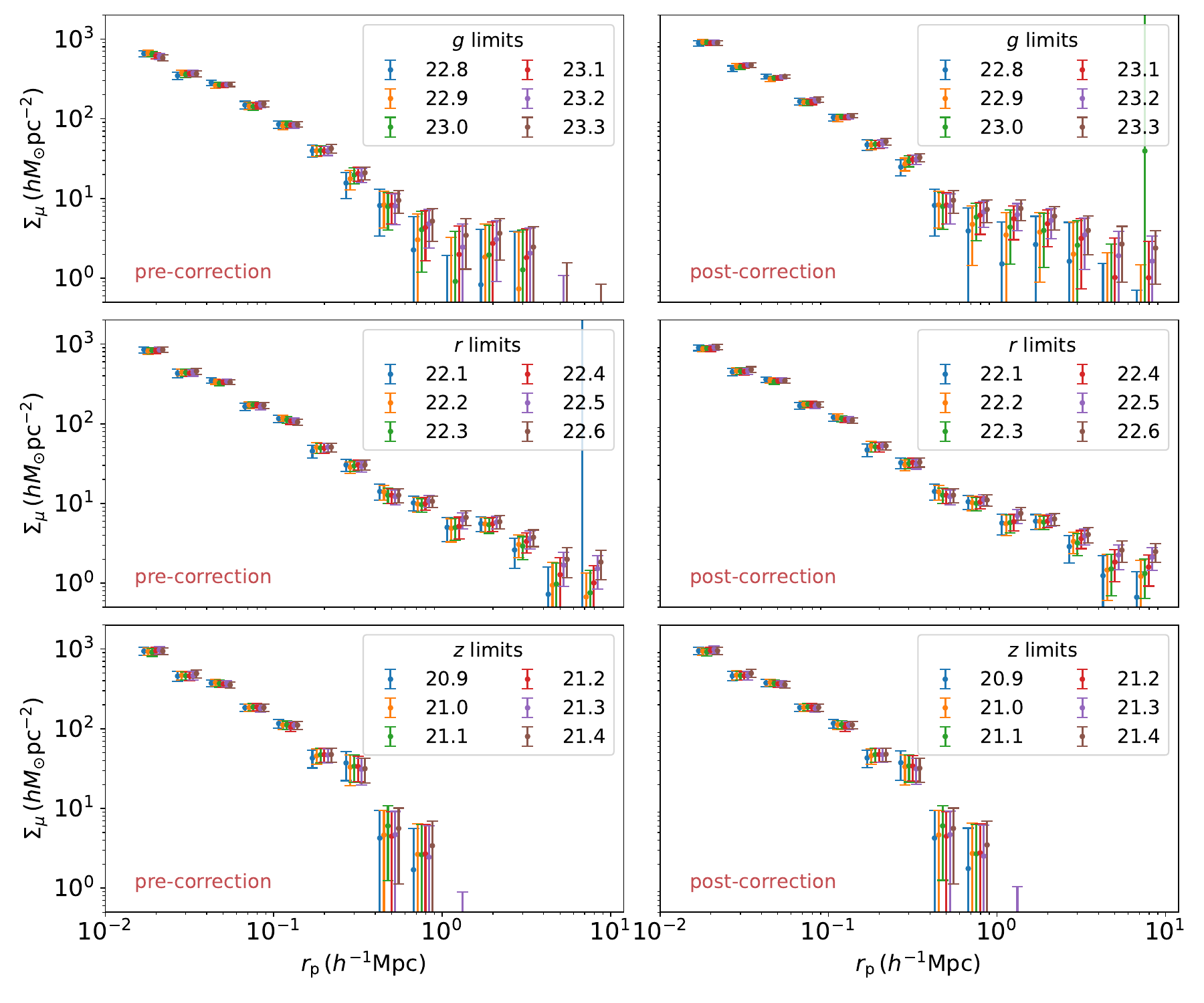}
    \caption{The measurements of $\Sigma_{\mu}$ for different magnitude limits in the $grz$ bands. Both pre- and post-dust correction results are depicted. The results are horizontally shifted for better visualization.}
    \label{fig:multi_limits}
\end{figure*}

\subsection{Constraining the Dust Attenuation}\label{sec:dust_chi2}
Now that we have established the form of the dust attenuation curves, we can directly model the $\delta M_{\lambda}^{\rm{obs}}(r_{\rm{p}})$ in the $grz$ bands using Equation \ref{eq:mag_p_dust} and constrain the distributions of dust attenuation $A_{\lambda}(r_{\rm{p}})$ around CMASS lens galaxies. For each $r_{\rm{p}}$ bin, we can express the dust-corrected $\Sigma_{\mu,\lambda}^{\rm{corr}}$ for the $grz$ bands as:
\begin{equation}
    \Sigma_{\mu,\lambda}^{\rm{corr}}=\frac{\Sigma_{\rm{crit}}}{2}G_{\lambda}^{-1}(\delta M_{\lambda}^{\rm{obs}}-A_{\lambda})\,,
\end{equation}
where $G^{-1}_{\lambda}$ denotes the inverse function of $G_{\lambda}$, and $A$ is parameterized using Equation \ref{eq:atte_curve} with $\delta$ set to $-5$. We compute $\Sigma_{\mu,\lambda}^{\rm{corr}}$ for each jackknife sub-sample and derive the mean value and covariance matrix of $\mathbf{\Sigma}_{\mu}^{\rm{corr}}=(\Sigma_{\mu,g}^{\rm{corr}},\Sigma_{\mu,r}^{\rm{corr}},\Sigma_{\mu,z}^{\rm{corr}})^T$. The $\chi^2$ is defined as:
\begin{equation}
    \chi^2=(\mathbf{\Sigma}_{\mu}^{{\rm{corr}}}-\Sigma_{\mu})^T\mathbf{C}^{-1}(\mathbf{\Sigma}_{\mu}^{{\rm{corr}}}-\Sigma_{\mu})\,,\label{eq:dust_chi2}
\end{equation}
where $\mathbf{C}^{-1}$ represents the inverse of the covariance matrix of $\mathbf{\Sigma}_{\mu}^{\rm{corr}}$, and $\Sigma_{\mu}$ is the intrinsic $\Sigma_{\mu}$ to be constrained. We employ the Markov Chain Monte Carlo (MCMC) sampler \texttt{emcee} \citep{2013PASP..125..306F} to conduct maximum likelihood analyses of $\Sigma_{\mu}$ and the amplitude of the attenuation curve.

The dust attenuation in various bands around CMASS lens galaxies is depicted in Figure \ref{fig:A_lambda}. The sharp decline in attenuation from the $g$ band to the $z$ band illustrates the steep nature of the attenuation curve. The attenuation decreases with $r_{\rm{p}}$ at a rate approximately proportional to $r_{\rm{p}}^{-1}$ and becomes flatter when reaching 0.5$\,h^{-1}$Mpc. These features may indicate the one-halo and two-halo distributions of the dust, where results with $r_{\rm{p}}<0.5\,h^{-1}$Mpc show the dust distribution within the halos, and those of $r_{\rm{p}}>0.5\,h^{-1}$Mpc denote the distributions outside the halos.

In the right panel of Figure \ref{fig:A_lambda}, we compare the $V$ band dust attenuation from our results to those from \citet{2010MNRAS.405.1025M}, where they measure the $V$ band dust attenuation around all galaxies with $17<i<21$. We observe a higher $A_V$ around CMASS galaxies, suggesting that more massive halos may contain more dust. With the current quality of data, we have not identified significant differences in the $r_{\rm{p}}$ dependence of $A_V$ between these two samples.

\subsection{Dust-corrected Magnification Results}\label{sec:dust_corrected}
With the distributions of dust attenuation in the $grz$ bands, we can now adjust for the dust attenuation in the magnification measurements and derive the intrinsic magnification signals around CMASS galaxies. We calculate the intrinsic $\delta M$ in each band using the best-fit dust attenuation:
\begin{equation}
    \delta M_{\lambda}(r_{\rm{p}})=\delta M_{\lambda}^{\rm{obs}}(r_{\rm{p}})-A_{\lambda}(r_{\rm{p}})\,.
\end{equation}
Subsequently, we compute the mean values and covariance matrices for $\delta M$, $\delta \mu$, and $\Sigma_{\mu}$ in each band, following the procedure outlined in Section \ref{sec:measurements}.  We present the results with corrections for dust attenuation in Figure \ref{fig:measure_post}. 

Furthermore, as discussed in Section \ref{sec:dust_chi2}, we incorporate constraints on $\Sigma_{\mu}$ in the fitting procedure, which tends to be more model-dependent, especially for error estimation. Moreover, deriving the covariance matrix across different $r_{\rm{p}}$ bins for $\Sigma_{\mu}$ is non-trivial and more complex. Nevertheless, we also illustrate the constraints on $\Sigma_{\mu}$ from the fitting procedure in Section \ref{sec:dust_chi2} in the right panel of Figure \ref{fig:measure_post}.

After correcting for dust attenuation, as illustrated in Figure \ref{fig:measure_post}, the differences in $\Sigma_{\mu}$ among different bands, as observed in Figure \ref{fig:measure_pre}, are eliminated. We notice a consistent $\Sigma_{\mu}$ across the $grz$ bands, along with the constraints from the dust model, validating the effectiveness of our dust correction. This also demonstrates that breaking the degeneracy between lensing magnification and dust attenuation with multi-band measurements is feasible.

To further validate our results, we present the outcomes for various magnitude limits in the $grz$ bands in Figure \ref{fig:multi_limits}, with the largest differences of 0.5 magnitudes brighter. We display both pre (left) and post-dust correction results (right). The dust attenuation curves constrained using the fiducial samples are applied to all other magnitude limits, as the dust attenuation curves should not depend on source samples. Although the samples comprise various sources with distinct $\delta \mu-\delta M$ relations and redshift distributions, the measurements $\Sigma_{\mu}$ across all samples remain consistent within $1\sigma$ in both pre- and post-dust correction analyses. Notably, post-dust correction results exhibit improved agreement among different magnitude limits, particularly evident in the smallest $r_{\rm{p}}$ bin of the $g$ bands. In this $r_{\rm{p}}$ bin, for pre-dust correction results, $\Sigma_{\mu}$ decreases with magnitude limits, while it agrees perfectly for post-dust correction results. These findings affirm the stability and reliability of our magnification measurements and dust attenuation corrections, validating that our results are not dominated by systematic errors. Although the use of total flux density, $\delta M$, as the measurement can reduce some systematics related to number density in image processing, the obscuration of background galaxies by CMASS galaxies and the biased estimation of fluxes in high-density regions remain concerns. Additionally, potential contamination in the source samples from galaxies with redshifts overlapping with CMASS could bias the magnification measurements. However, if the measurements were dominated by systematic errors, the results in Figure \ref{fig:multi_limits} would suggest that these errors, very likely magnitude-dependent, affect $\delta M$ in a way that results in consistent systematics on $\Sigma_{\mu}$ for different magnitude-limited samples after applying different $\delta \mu-\delta M$ relations, correcting with the same dust attenuation value, and dividing by different $\Sigma_{\rm{crit}}$. We believe this scenario is not very likely. Therefore, the check in Figure \ref{fig:multi_limits} proves that our magnification measurements at the depth of DECaLS are still not dominated by systematics, which might be a more pronounced challenge for future deeper surveys as they will extract much fainter galaxies.

\begin{table*}
    \caption{Essential details of the simulations employed for comparison with observations.}
    \centering
    \begin{tabular}{cccccccc}
    \toprule

    Simulation & Cosmology & type & $L$ & $m_{\rm{DM}}$ & $m_{\rm{gas}}$ &$\epsilon$&$S_8$\\
     & & & $h^{-1}$Mpc & $h^{-1}M_{\odot}$ & $h^{-1}M_{\odot}$&$h^{-1}$Mpc&\\
    \midrule
    {\texttt{CosmicGrowth}} & WMAP9 & DMO & 600 & $5.54\times10^8$&&0.01&0.785\\
    {\texttt{Jiutian}} & Planck18 & DMO & 1000 & $3.72\times10^8$&&0.004&0.825\\
    {\texttt{TNG300-1}} & Planck15 & Hydro & 205 & $3.98\times10^7$&$7.44\times10^6$ &0.001&0.828\\
    {\texttt{TNG300-1-Dark}} & Planck15 & DMO & 205 & $4.73\times10^7$& &0.001&0.828\\
    \bottomrule
    \end{tabular}
    \label{tab:simu}
\end{table*}

\begin{table*}
    \caption{Posterior PDFs of the parameters of the double power law SHMR models for WMAP9 and Planck18 cosmologies.}
    \centering
    \begin{tabular}{ccccccc}
    \toprule

    Cosmology&Simulation & $\log_{10}(M_0/h^{-1}M_{\odot})$ & $\alpha$ & $\beta$ & $\log_{10}(k/M_{\odot})$ & $\sigma$\\
    \midrule
    WMAP9 & {\texttt{CosmicGrowth}} & $11.624^{+0.010}_{-0.010}$ & $0.466^{+0.008}_{-0.008}$ & $2.513^{+0.034}_{-0.033}$ & $10.133^{+0.010}_{-0.010}$ & $0.192^{+0.004}_{-0.004}$ \\
    Planck18 & {\texttt{Jiutian}} & $11.681^{+0.011}_{-0.011}$ &$0.438^{+0.008}_{-0.007}$
            & $2.531^{+0.037}_{-0.035}$ & $10.134^{+0.011}_{-0.011}$ &$0.212^{+0.003}_{-0.003}$\\
    \bottomrule
    \end{tabular}
    \label{tab:SHMR}
\end{table*}

\begin{figure*}
    \plotone{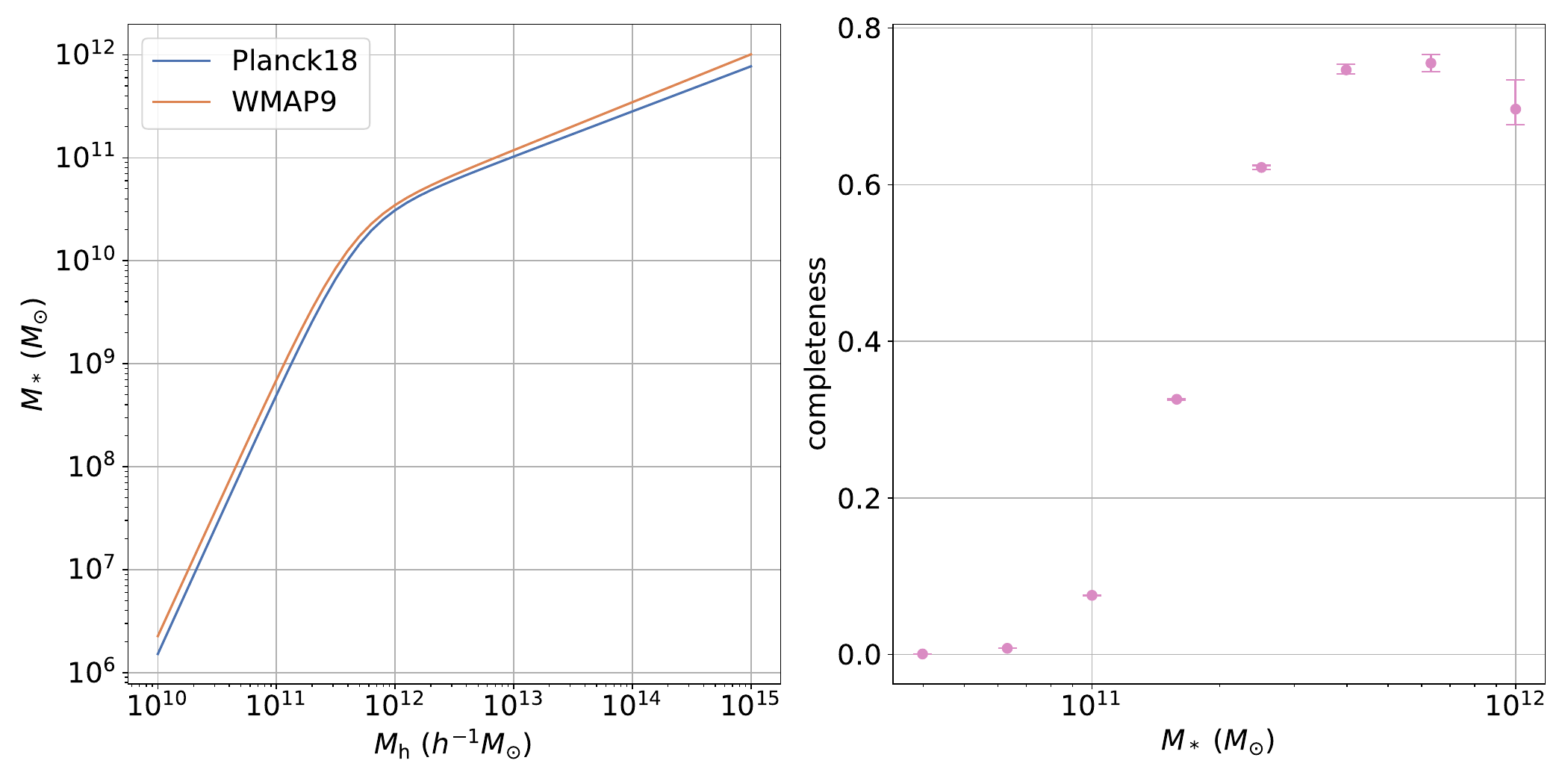}
    \caption{Left: The mean stellar-halo mass relations at $z\sim0.6$ in Planck18 and WMAP9 cosmologies, modeled based on the PAC measurements from \citet{2023ApJ...944..200X}. Right: Stellar mass completeness of the CMASS sample within $0.5<z<0.65$, derived by comparing the number density of CMASS galaxies to the GSMF from \citet{2022ApJ...939..104X} of the entire galaxy population.}
    \label{fig:SHMR}
\end{figure*}

\section{Simulation data}\label{sec:simulation}
In this section, we outline three simulations employing different cosmological and galaxy formation models, namely: \texttt{CosmicGrowth}, \texttt{Jiutian}, and \texttt{IllustrisTNG}, which are employed for comparison with our observational findings. The critical information regarding these simulations is outlined in Table \ref{tab:simu}.

\subsection{CosmicGrowth}
The \texttt{CosmicGrowth} simulation suite \citep{2019SCPMA..6219511J} comprises a grid of high-resolution N-body simulations conducted in various cosmology utilizing an adaptive parallel P$^3$M code \citep{2002ApJ...574..538J,2007ApJ...657..664J}. We utilize one of the $\Lambda$CDM simulations featuring the WMAP9 cosmological parameters: $\Omega_m = 0.268$, $\Omega_{\Lambda} = 0.732$, and $\sigma_8 = 0.831$ \citep{2013ApJS..208...19H}. The simulation box spans $600\ h^{-1}{\rm{Mpc}}$, comprising $3072^3$ dark matter particles with a softening length of $\epsilon = 0.01\ h^{-1}{\rm{Mpc}}$. Group identification employs the friends-of-friends (FOF) algorithm with a linking length set to 0.2 times the mean particle separation. Subsequently, halos undergo processing with HBT+ \citep{2012MNRAS.427.2437H,2018MNRAS.474..604H} to detect subhalos and track their evolutionary paths. We utilize catalog snapshots at redshifts approximately around $0.57$ to compare with magnification measurements around CMASS lens galaxies. Merger timescales for subhalos containing fewer than 20 particles, which might be unresolved, are assessed using the fitting formula in \citet{2008ApJ...675.1095J}, with those already merged into central subhalos discarded. The halo mass function \citep[see Figure 1]{2019SCPMA..6219511J} and subhalo mass function \citep[see Figure 4]{2022ApJ...925...31X} derived from the \texttt{CosmicGrowth} simulation exhibit robustness down to at least 20 particles ($\sim10^{10.0}h^{-1}M_{\odot}$), sufficient for our purposes.

\subsection{Jiutian}
The {\texttt{Jiutian}} suite comprises a series of N-body simulations developed to fulfill the scientific requirements of the Chinese Space Station Telescope (CSST) optical surveys \citep{2011SSPMA..41.1441Z,2019ApJ...883..203G}. We employ one of the high-resolution main runs based on the Planck18 cosmology \citep{2020A&A...641A...6P}, featuring $\Omega_m = 0.3111$, $\Omega_{\Lambda}=0.6889$, and $\sigma_8 = 0.8102$. This simulation box spans $1000\ h^{-1}{\rm{Mpc}}$ and consists of $6144^3$ dark matter particles with a softening length of $\epsilon=0.004h^{-1}$Mpc, simulated using the {\texttt{GADGET-3}} code \citep{2001NewA....6...79S,2005ApJ...633..589S}. Dark matter halos are identified using the FOF algorithm with a linking length set to 0.2 times the mean interparticle separation. Subsequently, these halos undergo further processing with the newly implemented {\texttt{HBT+}} code. Subhalos are defined with a minimum of 20 particles, and when a subhalo is no longer resolved, it retains track of its most-bound particle. Merger timescales for the unresolved subhalos are estimated using the fitting formula proposed by \citet{2008ApJ...675.1095J}, with those already merged into central subhalos being discarded. We utilize catalog snapshots at redshifts approximately around $z\sim0.59$ to compare with magnification measurements around CMASS lens galaxies.

\subsection{IllustrisTNG}
The \texttt{IllustrisTNG} simulations constitute a suite of magnetohydrodynamic cosmological simulations \citep{2018MNRAS.480.5113M, 2018MNRAS.475..624N, 2018MNRAS.477.1206N,2018MNRAS.475..676S,2018MNRAS.475..648P,2019ComAC...6....2N}. These simulations are performed using the moving mesh code {\texttt{AREPO}} \citep{2010MNRAS.401..791S} and incorporate various baryonic processes implemented as sub-grid physics \citep{2017MNRAS.465.3291W,2018MNRAS.473.4077P}. The cosmological parameters employed in the simulations are consistent with Planck15 results \citep{2016A&A...594A..13P}: $\Omega_m=0.3089$, $\Omega_{\Lambda}=0.6911$, $\Omega_b=0.0486$, $h=0.6774$, $\sigma_8=0.8159$, $n_s=0.9667$. Among the simulation suite, we utilize the run with the largest box, {\texttt{TNG300-1}}. The box size is $205h^{-1}{\rm{Mpc}}$, containing $2500^{3}$ dark matter particles and $2500^{3}$ gas cells, corresponding to mass resolutions of $m_{\rm{DM}}=3.98\times10^{7}h^{-1}M_{\odot}$ and $m_{\rm{gas}}=7.44\times10^{6}h^{-1}M_{\odot}$. The maximun softening length is $\epsilon=0.001h^{-1}$Mpc. Dark matter halos within the simulation were cataloged using FoF methods, while subhalos were cataloged using the {\texttt{SUBFIND}} algorithm \citep{2001MNRAS.328..726S}. The simulation output is stored in 100 snapshots for redshifts ranging from $z=20$ to $z=0$. We utilize the snapshot at $z=0.58$ to compare with magnification measurements around CMASS lens galaxies. 

As demonstrated in the following section, the galaxy-halo connection of the CMASS galaxies is constrained within DMO simulations. Consequently, we also utilize the corresponding DMO run \texttt{TNG300-1-Dark} to identify the (sub)halos of CMASS galaxies, subsequently aligning them with \texttt{TNG300-1}. \texttt{TNG300-1-Dark} adopts identical cosmological parameters, box size, and initial conditions as \texttt{TNG300-1}, featuring $2500^3$ dark matter particles with a mass resolution of $m_{\rm{DM}}=4.73\times10^{7}h^{-1}M_{\odot}$.

\section{Model Predictions of Lensing Magnification around CMASS Galaxies}\label{sec:model_pred}
In this section, leveraging the high-resolution simulations, the precise galaxy-halo connection for CMASS galaxies obtained through the Photometric objects Around Cosmic webs (PAC) method \citep{2022ApJ...925...31X,2023ApJ...944..200X}, and the accurate 2D particle-particle-particle-mesh (P$^3$M) algorithm \texttt{P3MLens} \citep{2021ApJ...915...75X} for ray-tracing, we generate model predictions of lensing magnification around CMASS galaxies in various cosmological and galaxy formation models, subsequently comparing them to observations.

\begin{figure}
    \plotone{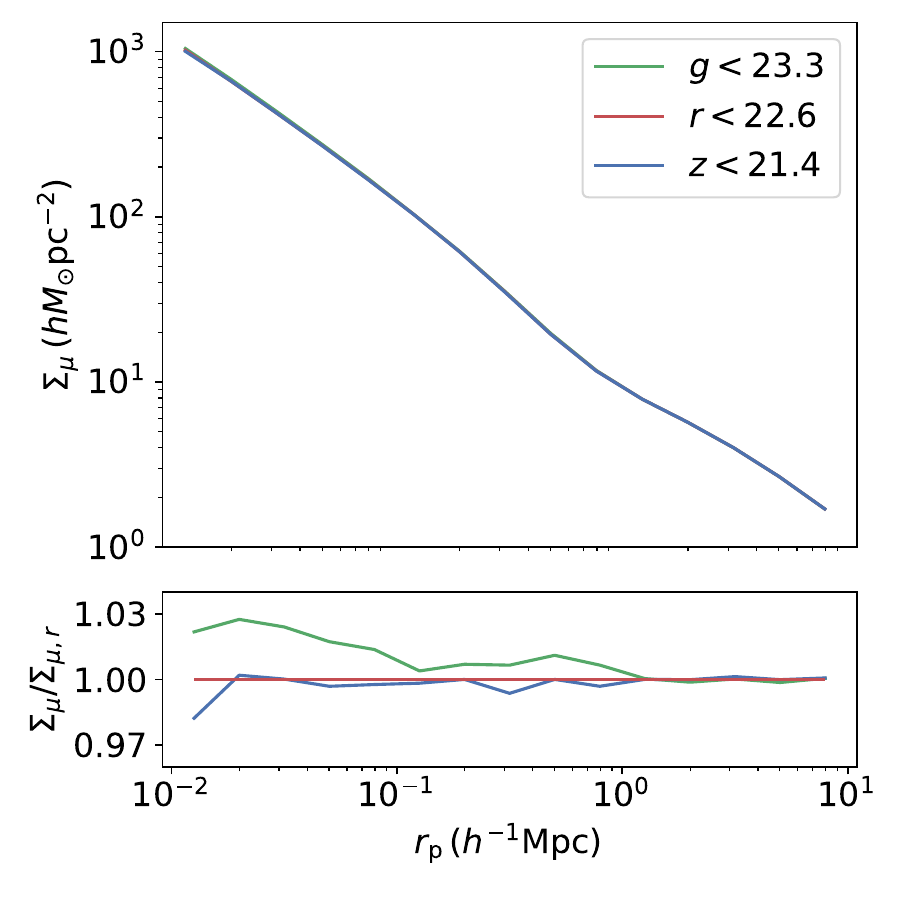}
    \caption{Model predictions of $\Sigma_{\mu}$ from {\texttt{Jiutian}} for three fiducial source samples with distinct redshift distributions.}
    \label{fig:sigma_grz}
\end{figure} 

\begin{figure}
    \plotone{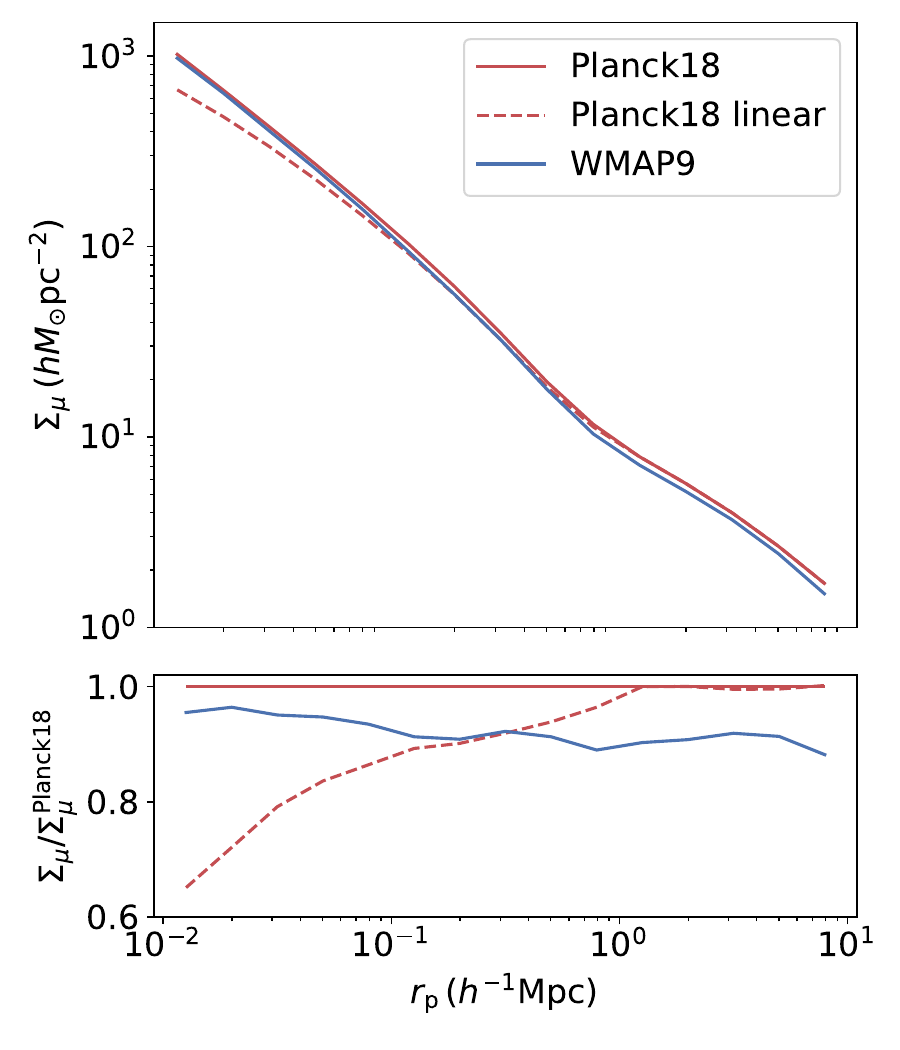}
    \caption{DMO model predictions of $\Sigma_{\mu}$ in the Planck18 and WMAP9 cosmologies from {\texttt{Jiutian}} and {\texttt{CosmicGrowth}} are presented. Additionally, the linear approximation of $\Sigma_{\mu}$ in the weak lensing regime for Planck18 is depicted, which corresponds to the matter surface density $\Sigma$.}
    \label{fig:sigma_cosmo}
\end{figure} 

\subsection{Photometric objects Around Cosmic webs (PAC)}
A precise comprehension of the galaxy-halo connection for CMASS lens galaxies is vital for faithfully replicating the magnification signal in simulations. We employ the galaxy-halo connection for CMASS galaxies derived by \citet{2023ApJ...944..200X}, which utilized the subhalo abundance matching (SHAM) technique \citep{2006MNRAS.371..537W, 2006MNRAS.371.1173V, 2010ApJ...717..379B, 2010MNRAS.402.1796W, 2010MNRAS.404.1111G,2013MNRAS.428.3121M,2023ApJ...944..200X} to model measurements from the PAC method \citep{2022ApJ...925...31X}. The PAC method integrates cosmological spectroscopic and photometric surveys, offering abundant information for studying the galaxy-halo connection across the entire galaxy population and specific spectroscopic tracers. PAC can measure the excess surface density $\bar{n}_2w_{\rm{p}}(r_{\rm{p}})$ of photometric objects with specific physical properties, such as stellar mass, around spectroscopic objects via angular correlations. Utilizing the CMASS spectroscopic sample and the DECaLS photometric sample, \citet{2023ApJ...944..200X} measured 33 $\bar{n}_2w_{\rm{p}}(r_{\rm{p}})$ across various stellar mass bins of both spectroscopic and photometric samples, down to $10^{9.7}M_{\odot}$ at $z\sim0.6$. Subsequently, they applied SHAM within the \texttt{CosmicGrowth} simulation to model these measurements, achieving precise ($<1\%$) constraints on the stellar-halo mass relation (SHMR) for the entire galaxy population and the galaxy-halo connection specifically for the CMASS sample.

The galaxy-halo connection for the CMASS sample can be established through the combination of the SHMR and stellar mass completeness. For \texttt{CosmicGrowth}, we utilize the outcomes derived from the double power law model of the SHMR presented in \citet{2023ApJ...944..200X}, in which the SHMR is expressed by:
\begin{equation}
    M_{*} = \left[\frac{2k}{(M_{{\rm{acc}}}/{M_0})^{-\alpha}+(M_{{\rm{acc}}}/{M_0})^{-\beta}}\right]\,.\label{eq:8}
\end{equation}
Here, $M_{{\rm{acc}}}$ is defined as the viral mass $M_{{\rm{vir}}}$ of the halo at the time when the galaxy was last the central dominant object. The fitting formula in \citet{1998ApJ...495...80B} is employed to determine $M_{{\rm{vir}}}$. The scatter in $\log(M_*)$ at a given $M_{{\rm{acc}}}$ is described by a Gaussian function of width $\sigma$. For \texttt{Jiutian}, we re-model the PAC measurements from \citet{2023ApJ...944..200X} using SHAM with the double power law form of the SHMR to derive results under the Planck18 cosmology, which have also been used in \citet{2024arXiv240111997Z} and \citet{PAC_VII}. Regarding \texttt{TNG300-1}, its box size is inadequate for accurately modeling $\bar{n}_2w_{\rm{p}}(r_{\rm{p}})$ up to $10h^{-1}$Mpc. Therefore, we directly apply the SHMR from \texttt{Jiutian} to \texttt{TNG300-1-Dark}, and subsequently match the (sub)halos hosting the CMASS galaxies to \texttt{TNG300-1}. In the above process, we overlook the disparity between Planck18 and Planck15 cosmologies, a decision we find reasonable given the minor distinction, which will also be examined later. The constrained SHMRs in both Planck18 and WMAP9 cosmologies are presented in Table \ref{tab:SHMR} and illustrated in the left panel of Figure \ref{fig:SHMR}. 

In addition to the SHMR, establishing the galaxy-halo connection for CMASS requires knowledge of the stellar mass completeness of the CMASS sample \citep{2016MNRAS.457.4021L,2018ApJ...858...30G,2023ApJ...944..200X}. \citet{2023ApJ...944..200X} computed the stellar mass completeness for CMASS galaxies within $0.5<z<0.7$ and four narrower redshift bins by comparing the number density of CMASS galaxies to the galaxy stellar mass function (GSMF) from \citet{2022ApJ...939..104X} representing the entire galaxy population. As we utilize a different redshift range, $0.5<z<0.65$, for CMASS in this study, we recompute the stellar mass completeness for CMASS galaxies within this range following \citet{2023ApJ...944..200X}. The stellar mass completeness of the CMASS sample for $0.5<z<0.65$ is displayed in the right panel of Figure \ref{fig:SHMR}. After populating (sub)halos with galaxies according to the SHMR, the galaxy-halo connection of CMASS galaxies can be established by randomly reducing the number density in each stellar mass bin based on the stellar mass completeness illustrated in the right panel of Figure \ref{fig:SHMR}.

\begin{figure}
    \plotone{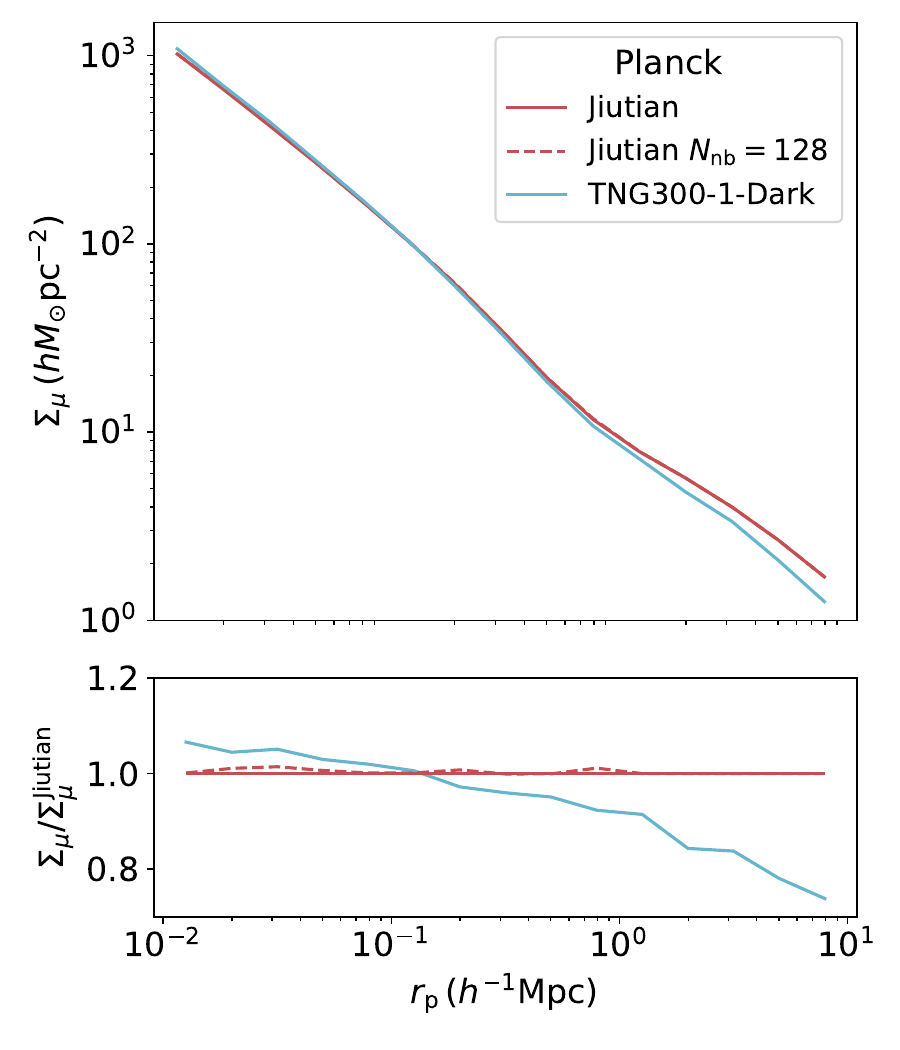}
    \caption{DMO model predictions of $\Sigma_{\mu}$ in the Planck cosmology from simulations with varying mass resolutions and boxsizes, comparing {\texttt{Jiutian}} and {\texttt{TNG300-1-Dark}}. {\texttt{TNG300-1-Dark}} boasts a mass resolution approximately 8 times greater than {\texttt{Jiutian}}, albeit with a volume approximately 110 times smaller. Additionally, the results of {\texttt{Jiutian}} with a smaller soften length ($N_{\rm{nb}}=128$) in ray-tracing using {\texttt{P3MLens}} are also displayed.}
    \label{fig:sigma_res}
\end{figure} 

\begin{figure*}
    \plottwo{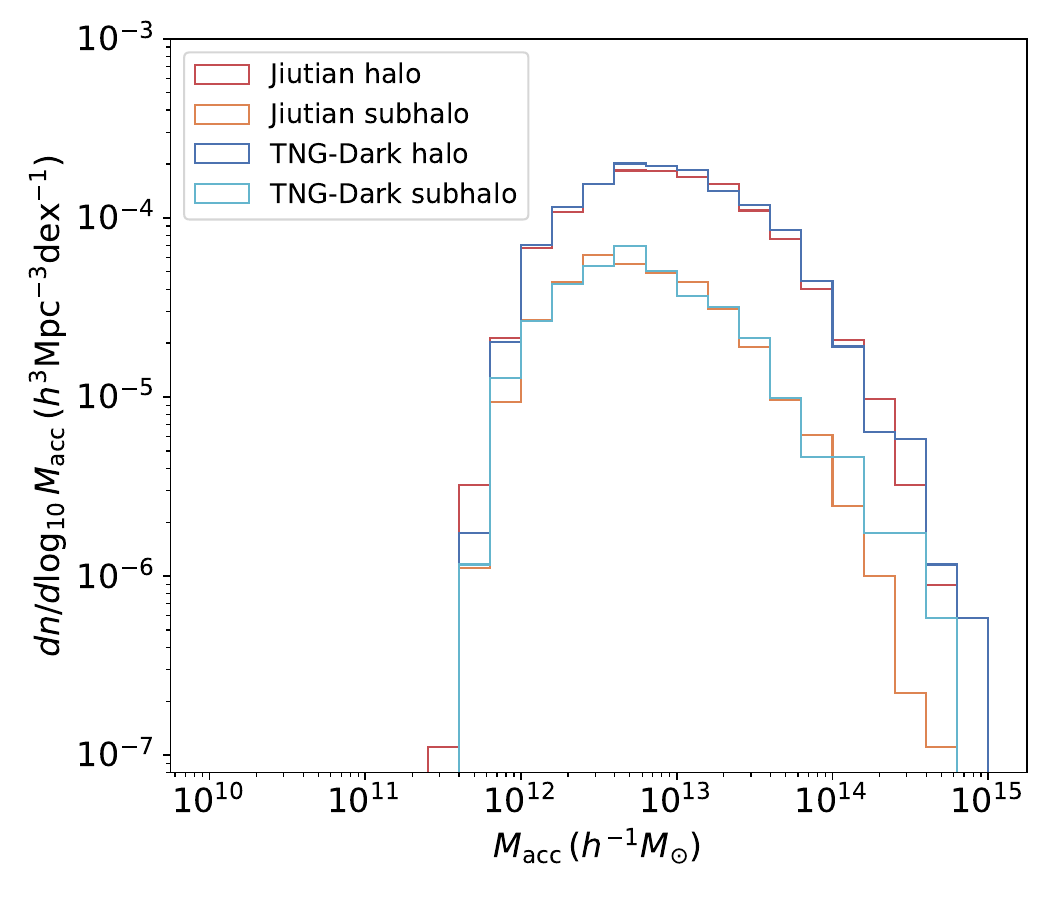}{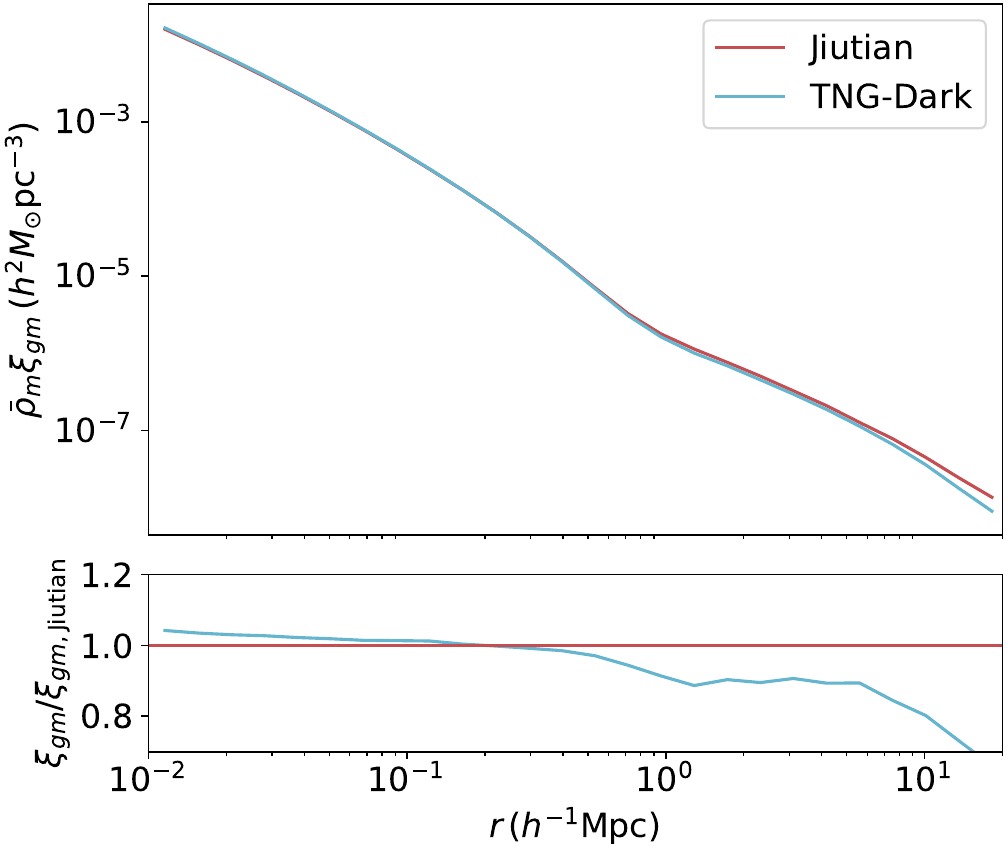}
    \caption{Left: The mass distributions of halos and subhalos hosting CMASS galaxies in the {\texttt{Jiutian}} and {\texttt{TNG300-1-Dark}} simulations with Planck cosmology. Right: The 3D matter distribution $\bar{\rho}_m\xi_{gm}$ for halos in the {\texttt{Jiutian}} and {\texttt{TNG300-1-Dark}} simulations. The mass distribution of the halos is controlled to match exactly that of the halos hosting CMASS galaxies in the {\texttt{TNG300-1-Dark}} simulation.}
    \label{fig:halo_mass}
\end{figure*} 

\subsection{P3MLens}
After pinpointing the (sub)halos that host CMASS galaxies in simulations, we have acquired knowledge about the matter distributions surrounding them. To generate the model prediction for magnification, we opt to conduct ray-tracing around these (sub)halos utilizing the thin lens approximation \citep{2000ApJ...530..547J}. However, since we are delving into the innermost regions of the massive halos, the particle-mesh (PM) algorithm typically employed for gravitational lensing studies \citep{2000ApJ...530..547J,2016A&C....17...73P} may prove inefficient in achieving the required accuracy. This limitation arises because the PM algorithm is primarily accurate for long-range forces on scales larger than a few grid sizes. In extremely dense regions where the mean particle separation is very short, setting the grid size to be very small is necessary to achieve highly accurate modeling, which can potentially degrade the PM algorithm into a brute-force method. 

To effectively model the magnification signal, we utilize the recently developed 2D P$^3$M algorithm \texttt{P3MLens}\footnote{https://github.com/kunxusjtu/P3MLens} \citep{2021ApJ...915...75X}. Equipped with optimized Green's functions and adaptive soften length, \texttt{P3MLens} attains an average force accuracy smaller than 0.1\% across all scales and achieves percent-level accuracy for lens quantities at 0.1 scale radius $r_{\rm{s}}$ for massive halos in simulations with resolutions similar to those used in this study. We discover that \texttt{P3MLens} is highly suitable for analyzing lensing signals around CMASS galaxies down to small scales and intend to employ it for comparison with observational data.

\begin{figure}
    \plotone{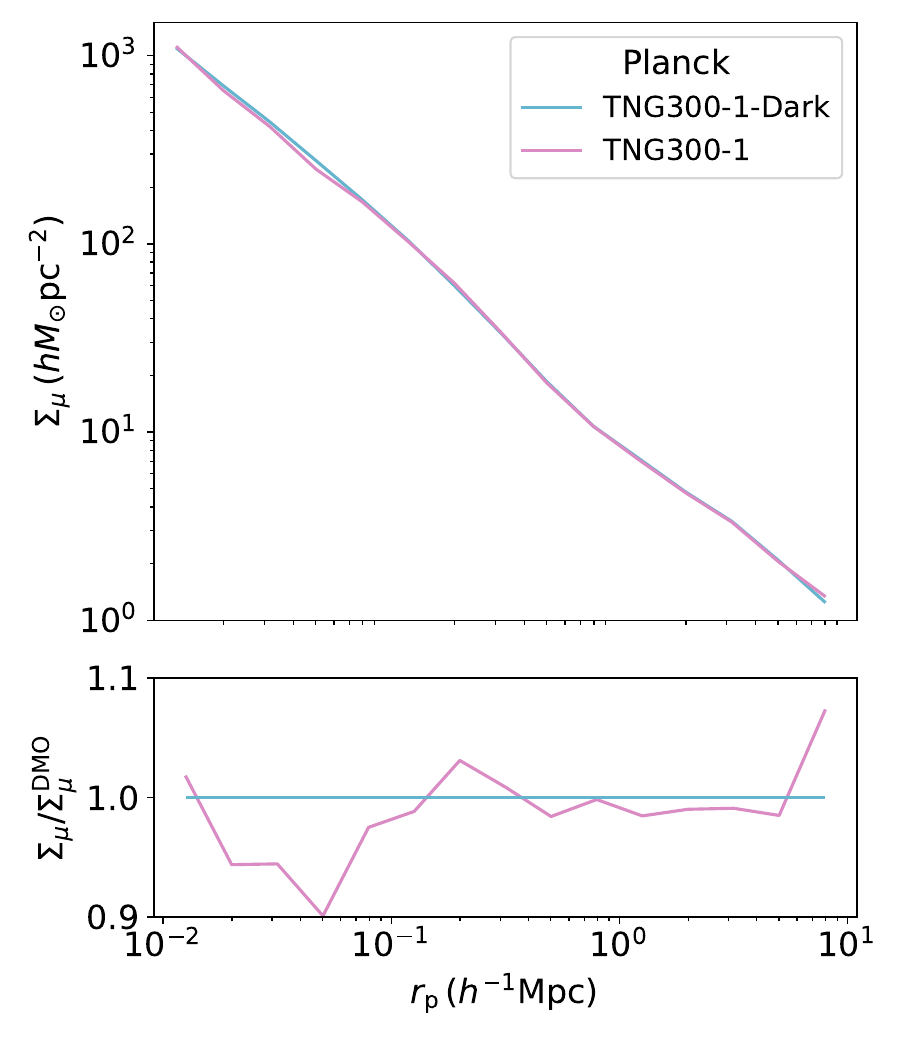}
    \caption{Model predictions of $\Sigma_{\mu}$ in the hydrodynamic simulation \texttt{TNG300-1} and its DMO companion \texttt{TNG300-1-Dark}.}
    \label{fig:sigma_tng}
\end{figure} 

\begin{figure*}
    \plotone{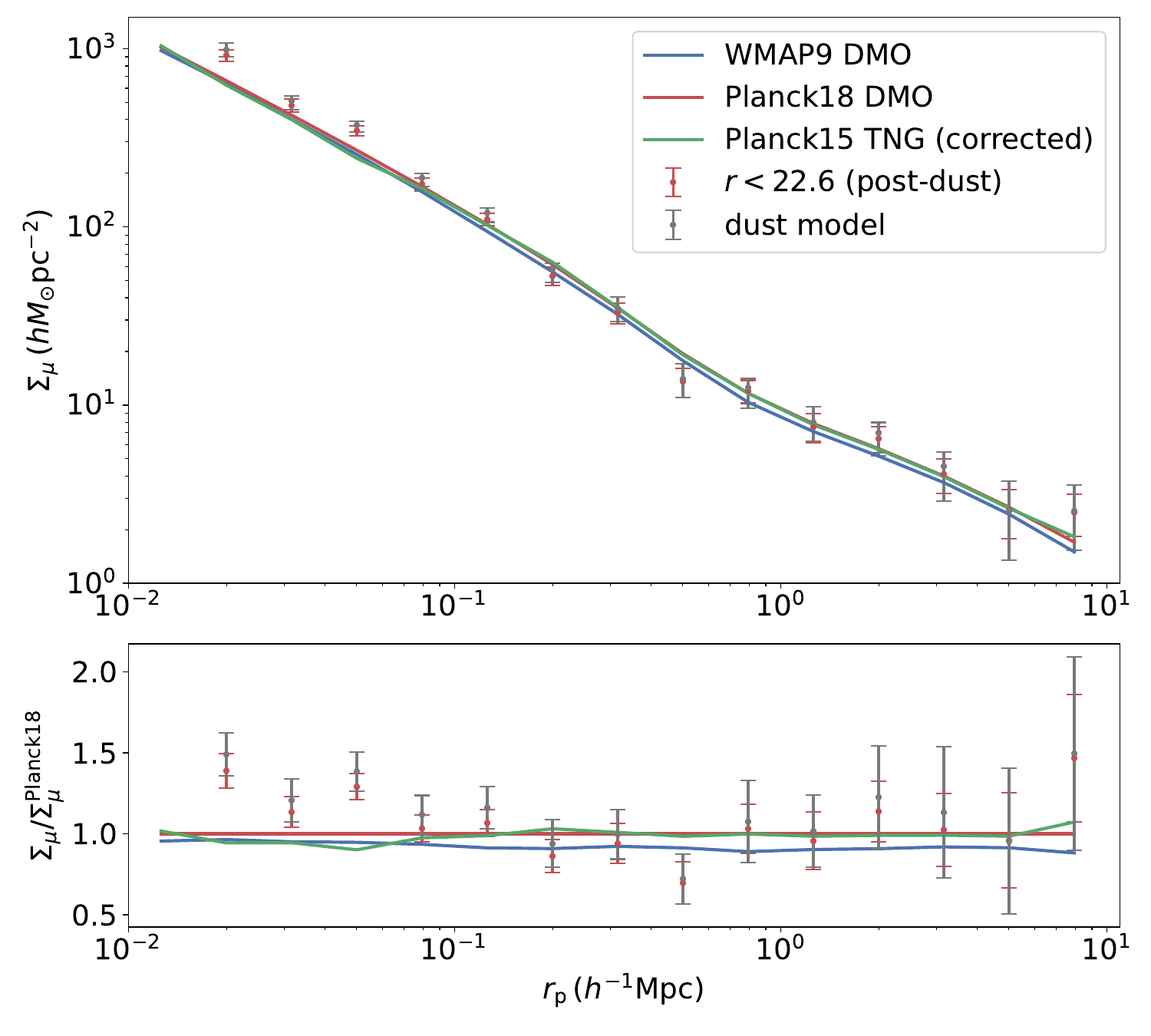}
    \caption{Comparing the $\Sigma_{\mu}$ around CMASS lens galaxies in observations to model predictions from simulations with different cosmological and galaxy formation models. The results from the $r<22.6$ source sample after dust corrections with the best-fit $A_{\rm{r}}^{0.56}$ (red dots) and the combined constraints from three fiducial $grz$ limited samples with dust attenuation models (grey dots) are displayed for observations. Two DMO models with WMAP9 (blue lines) and Planck18 (red lines) cosmologies and one hydrodynamic model from \texttt{TNG300-1} after correcting for the limited box problems (green lines) are represented. All results are compared with the Planck18 DMO model in the bottom panel.}
     \label{fig:sigma_compare}
\end{figure*} 

\begin{figure}
    \plotone{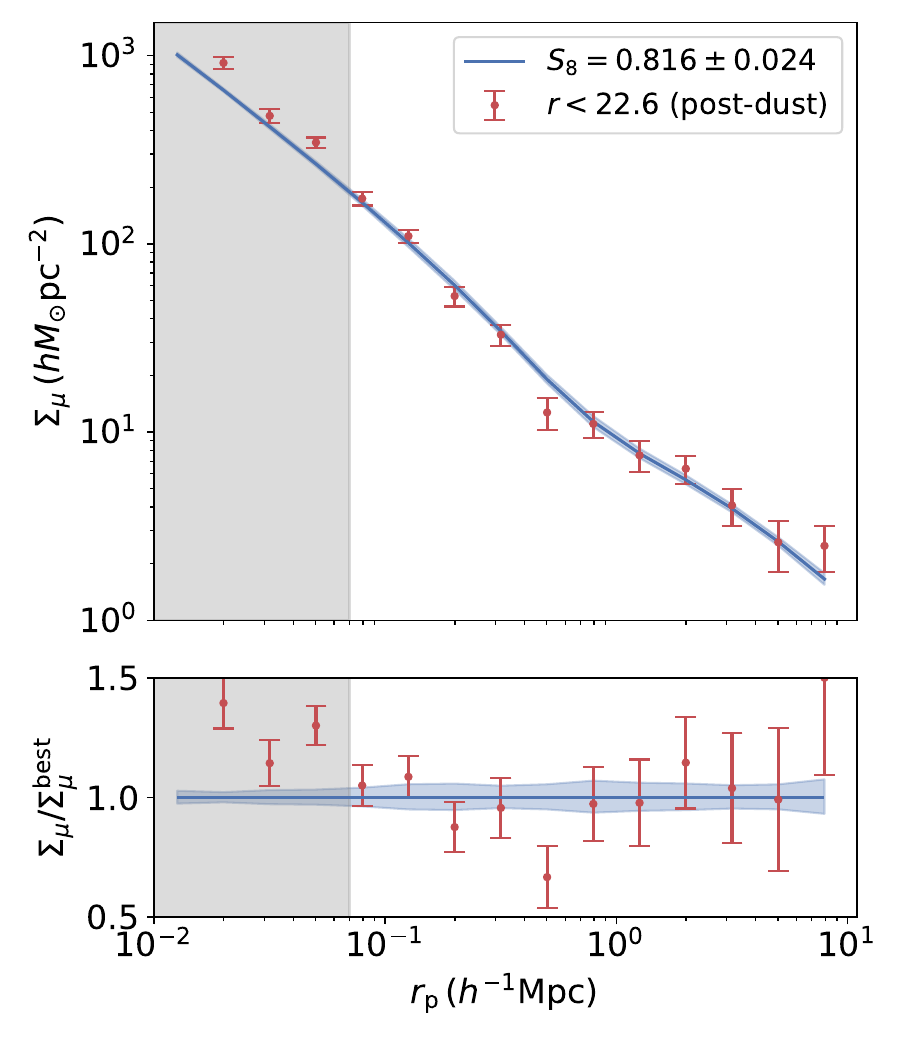}
    \caption{The best-fit $\Sigma_{\mu}$ and the 1$\sigma$ confidence level for the measurements from the $r<22.6$ source samples with $S_8=0.816\pm0.024$.}
    \label{fig:sigma_fitting}
\end{figure} 

\subsection{Magnification around CMASS Galaxies in Simulations}\label{sec:simu_lens}
Based on the SHMR and stellar mass completeness depicted in Figure \ref{fig:SHMR}, we identify (sub)halos hosting CMASS galaxies in the simulations {\texttt{CosmicGrowth}}, {\texttt{Jiutian}}, and {\texttt{TNG300-1-Dark}}. For each simulation, we designate the $Z$-direction as the line-of-sight direction and construct a thin lens centered on each (sub)halo with a width of $\Delta Z=100h^{-1}$Mpc. Subsequently, we randomly select 1000 shot points within each $r_{\rm{p}}$ bin for each (sub)halo and compute the deflection angles using \texttt{P3MLens}. We choose a softening length of $N_{\rm{nb}}=512$ in \texttt{P3MLens}, which means that for each particle, the softening length is chosen as the distance to its 512th neighboring particles. Next, we calculate the magnification $\mu$ for each point by randomly selecting redshift pairs from the lens and source samples in the observation. Finally, we average $\mu$ within each $r_{\rm{p}}$ bin and convert it to $\Sigma_{\mu}$ using the mean redshifts of the lens and source samples to maintain consistency with the observations.

We initially validate the assumptions outlined in Section \ref{sec:dust_theory} regarding the negligible redshift dependence of $\Sigma_{\mu}$ for our source samples. We illustrate the predicted $\Sigma_{\mu}$ from {\texttt{Jiutian}} for three different redshift distributions of our source samples in Figure \ref{fig:sigma_grz}. It is evident that the $\Sigma_{\mu}$ of the $g<23.3$ sample exhibits a larger deviation compared to the other two, primarily due to its more different mean redshift. This difference amplifies with the magnification signal, reaching up to $3\%$ in the innermost region. This analysis supports our assumption that the redshift dependence of $\Sigma_{\mu}$ is very weak for our source samples. In the subsequent analysis, we employ the $\Sigma_{\mu}$ models obtained from the $r<22.6$ sample as our fiducial simulation results.

In Figure \ref{fig:sigma_cosmo}, we present the model predictions of $\Sigma_{\mu}$ for two distinct cosmologies, Planck18 and WMAP9, utilizing {\texttt{Jiutian}} and \texttt{CosmicGrowth}. Notably, with a 5\% lower $S_8=\sigma_8(\Omega_m/0.3)^{0.5}$ in WMAP9 compared to Planck18, we observe an approximately $10\%$ reduction in $\Sigma_{\mu}$ in WMAP9, indicating a strong correlation between the magnification signal and $S_8$. Furthermore, to examine the non-linear effects in magnification, we contrast $\Sigma_{\mu}$ with its linear approximation $\Sigma$ in the Planck18 cosmology. We discover that non-linear effects become noticeable starting from $1\,h^{-1}$Mpc for massive (sub)halos hosting CMASS galaxies and become increasingly significant at smaller scales. By $20\,h^{-1}$kpc, the linear approximation of $\Sigma_{\mu}$ diminishes to only $70\%$ of the true values.

Before investigating the baryon effects on magnification using the hydrodynamic simulation \texttt{TNG300-1}, we perform a convergence test by comparing \texttt{Jiutian} to its DMO companion, \texttt{TNG300-1-Dark}. \texttt{TNG300-1-Dark} has a resolution approximately 8 times higher than \texttt{Jiutian}, but with a volume roughly 110 times smaller. In Figure \ref{fig:sigma_res}, we observe differences between {\texttt{TNG300-1-Dark}} and {\texttt{Jiutian}}, within $5\%$ at small scales ($r_{\rm{p}}<0.5h^{-1}$ Mpc) but increasing at larger scales, reaching $25\%$ at $10h^{-1}$ Mpc. We first check if this difference originates from the configuration used in \texttt{P3MLens} by employing a smaller softening length of $N_{\rm{nb}}=128$ in \texttt{Jiutian}. In Figure \ref{fig:sigma_res}, we observe no changes in $\Sigma_{\mu}$ after lowering the softening length. Therefore, the difference is from the simulations themselves. We suspect that the small box size of \texttt{TNG300-1-Dark} might be the issue, as both simulations have sufficient resolution for massive (sub)halos hosting CMASS galaxies. The low $k$ cut-off on the matter power spectrum due to the box size and the limited number of modes for low $k$ can introduce systematic and statistical errors in the matter distribution. Subsequently, this can affect the halo assembly history, resulting in different halo mass functions and halo structures \citep{2009ApJ...707..354Z}. For a clearer comparison, in the left panel of Figure \ref{fig:halo_mass}, we depict the mass distributions of halos and subhalos hosting CMASS galaxies in \texttt{Jiutian} and \texttt{TNG300-1-Dark}. We observe similar halo mass distributions with mean halo masses of $10^{13.28}h^{-1}M_{\odot}$ and $10^{13.29}h^{-1}M_{\odot}$ for \texttt{Jiutian} and \texttt{TNG300-1-Dark}, respectively. However, the mean subhalo mass in \texttt{TNG300-1-Dark} ($10^{13.18}h^{-1}M_{\odot}$) is $0.1$ dex higher than in \texttt{Jiutian} ($10^{13.08}h^{-1}M_{\odot}$). Moreover, in the right panel, to investigate the halo structure, after controlling the halo mass distributions to exactly match those of the halos hosting CMASS galaxies in \texttt{TNG300-1-Dark}, we compare the 3D matter distribution $\bar{\rho}_m\xi_{gm}$ around halos in \texttt{Jiutian} and \texttt{TNG300-1-Dark}, where $\bar{\rho}_m$ is the mean matter density of the Universe. We find that halos in \texttt{TNG300-1-Dark} have a slightly steeper profile than those in \texttt{Jiutian}. We believe that these two effects can account for the 5\% difference in $\Sigma_{\mu}$ at small scales. Furthermore, the larger difference in $\xi_{gm}$ at large scales, due to the small box size of \texttt{TNG300-1-Dark}, can also explain the 25\% difference in $\Sigma_{\mu}$.

While \texttt{TNG300-1} may also be affected by limitations due to a finite box size, comparing the magnification signal to its DMO companion, \texttt{TNG300-1-Dark}, can still accurately reveal the influence of baryonic effects. This is because they share the same initial conditions, and any deviations should be identical for them. To derive model predictions from \texttt{TNG300-1}, we match the (sub)halos hosting CMASS galaxies identified in \texttt{TNG300-1-Dark} to \texttt{TNG300-1}. This matching process is facilitated by \texttt{LHaloTree} \citep{2015A&C....13...12N}, and it is bidirectional. For each match, the best subhalo candidate is selected based on the largest number of matching DM particles. Only matches where the candidates agree in both directions (\texttt{TNG300-1}$\leftrightarrow$\texttt{TNG300-1-Dark}) are retained. We observe that all halos successfully find a match, but half of the subhalos are lost. This occurs because the subhalos have already merged with their central halos in \texttt{TNG300-1}, owing to differences in dynamic friction and merger timescales when accounting for galaxy formation. To achieve a more realistic matching, for subhalos, we match them at the time when they reach their historical maximum mass in \texttt{TNG300-1-Dark}, and then trace them in \texttt{TNG300-1} until $z=0.58$. As a result, we are utilizing their host halos for the subhalos lost in \texttt{TNG300-1}.

In Figure \ref{fig:sigma_tng}, we compare the model predictions of $\Sigma_{\mu}$ for CMASS galaxies in \texttt{TNG300-1} and \texttt{TNG300-1-Dark}. We find that \texttt{TNG300-1} exhibits a lower $\Sigma_{\mu}$ at $r_{\rm{p}}<100h^{-1}$ kpc, with a maximum difference of $10\%$ at $50h^{-1}$ kpc. This suppression is due to the baryon feedback processes implemented in the galaxy formation models of \texttt{TNG300-1} \citep{2018MNRAS.475..676S}. At larger scales, the $\Sigma_{\mu}$ from \texttt{TNG300-1} and \texttt{TNG300-1-Dark} show little difference.

Our analysis underscores the influence of both cosmological and galaxy formation models on $\Sigma_{\mu}$. It emphasizes the necessity of considering both factors when comparing model predictions with observational data.

\subsection{Comparing with Observations}
In Figure \ref{fig:sigma_compare}, we compare the measured $\Sigma_{\mu}$ around CMASS lens galaxies to the model predictions. The results from the $r<22.6$ source samples, which have the most accurate measurements among the three bands and are corrected for dust using the best-fit $A_{r}^{0.56}$, are displayed for observations (red dots). Additionally, we present the combined constraints from all three bands with the dust attenuation model (grey dots). For the model predictions, we present results from two DMO models in WMAP9 (blue lines) and Planck18 (red lines) cosmologies derived from \texttt{CosmicGrowth} and \texttt{Jiutian}. Additionally, we showcase the hydrodynamic model from \texttt{TNG300-1} (green lines). To address concerns related to the limited box size, the outcomes of \texttt{TNG300-1} are normalized by the $\Sigma_{\mu}$ ratio between \texttt{Jiutian} and \texttt{TNG300-1-Dark}, as illustrated in Figure \ref{fig:sigma_tng}. Consequently, the disparity between \texttt{TNG300-1} and \texttt{Jiutian} in Figure \ref{fig:sigma_compare} solely reflects the influence of baryon effects.

At $r_{\rm{p}}>70h^{-1}$kpc, we observe agreement between the observations and the model predictions derived from both the WMAP9 and Planck18 cosmologies, given the current level of accuracy. At these scales, the baryon effects from the \texttt{TNG300-1} model do not impact the results. Assuming a linear correlation between $\Sigma_{\mu}$ and $S_8$, we obtain a constraint of $S_8=0.816\pm0.024$ from the dust-corrected measurements of the $r<22.6$ source samples. The best-fit $\Sigma_{\mu}$ is dispalyed in Figure \ref{fig:sigma_fitting}. We opt not to utilize the observational data from the combined constraints of the dust model due to the complexity and non-trivial nature of obtaining the covariance matrix. We emphasize that this is only a {\it rough estimation} of $S_8$, indicating that our results are not in tension with those from the cosmic microwave background. More robust cosmological constraint would require emulators capable of simultaneously constraining cosmology and galaxy-halo connections through modeling both magnification and PAC measurements, while also accounting for uncertainties such as those arising from photometric redshifts.

At smaller scales, $r_{\rm{p}}<70h^{-1}$kpc, we observe that the measured $\Sigma_\mu$ is notably higher than the DMO model of Planck18 cosmology from \texttt{Jiutian}, with a $2.8\sigma$ difference when compared with the $r<22.6$ results. Moreover, including the \texttt{TNG300-1} galaxy formation models further increases the discrepancy, reaching $3.6\sigma$. This discrepancy may indicate an incomplete understanding of either the nature of dark matter or galaxy formation processes. We will discuss it further in the next section.

The accurate measurements of surface density across large scales illustrate the potential of lensing magnification for studying cosmology and galaxy formation.

\section{Discussion} \label{sec:discussion}
In this section, we delve into two intriguing findings from this study: the steep slope of the attenuation curves in the CGM of CMASS galaxies, and the high mass density observed in the innermost regions of (sub)halos.

\subsection{Steep Slop of the Attenuation Curves}
In Figure \ref{fig:fit_dust_two}, we observe that the attenuation curves measured in the CGM of massive CMASS galaxies exhibit a much steeper slope than those in starburst galaxies \citep{2000ApJ...533..682C}, reaching $\delta=-5$ when fitted with the modified \citet{2000ApJ...533..682C} attenuation curve (Equation \ref{eq:atte_curve}).  This may indicate that the dust component in the CGM of massive galaxies is different from that in starburst galaxies. To date, there are only a few direct constraints on dust reddening and attenuation curves in the CGM of galaxies. \citet{2010MNRAS.405.1025M} measured the reddening of background QSOs around galaxies with $17<i<21$ at $z\sim0.36$. These lens galaxies have much lower mean stellar mass than those in our studies. They found that the reddening in the CGM of those galaxies can be well described by the interstellar extinction law from \citet{1994ApJ...422..158O}, derived by analyzing the extinction of 22 Milky Way stars. However, \citet{2015ApJ...813....7P} measured the reddening in the CGM of nearby galaxies ($z\sim0.05$) and found an excess in the reddening of $u-g$ compared to the Small Magellanic Cloud (SMC) bar extinction curve \citep{2001ApJ...548..296W}. They also found that the extinction curve does not change much for stellar masses ranging from $6\times10^{9}M_{\odot}$ to $6\times10^{10}M_{\odot}$. Similarly, \citet{2012ApJ...754..116M} measured the reddening in MgII clouds at $z=1\sim2$ and also found an excess of reddening at $\lambda<0.2\,\mu$m compared to the SMC extinction curve.

The escalation of dust attenuation towards shorter wavelengths may hint at the presence of small grains within the CGM of galaxies. Nevertheless, the generation of dust particles smaller than $a\sim0.01\,\mu$m within the CGM persists as a challenge for contemporary theories \citep{1991ApJ...381..137F,1998MNRAS.300.1006D,2005MNRAS.358..379B,2017MNRAS.469..870H,2018MNRAS.474..604H,2019MNRAS.487..961H}. Dust may enter the CGM via galactic outflows propelled by supernovae and active galactic nuclei (AGN), or through radiation pressure. Several factors contribute to the impediment of transporting small grains into the CGM, including their limited time for shattering within the interstellar medium (ISM) before transport \citep{2017MNRAS.469..870H}, inefficiencies in responding to radiation pressure \citep{1998MNRAS.300.1006D,2019MNRAS.487..961H}, and deceleration due to gas drag in denser regions proximal to galaxies \citep{2005MNRAS.358..379B}. Consequently, current models typically predict that the typical size of grains injected from a galaxy into the CGM is around $a\sim0.1\,\mu$m. In recent developments, the proposal of dust shattering within the turbulence of the CGM has emerged as a promising mechanism to generate small grains and account for observations \citep{2021MNRAS.505.1794H,2024MNRAS.528.5008O,2024arXiv240500305H}, though a more sophisticated model is deemed necessary. Our measurements, along with their prospective expansion through next-generation cosmological surveys, hold the potential to enhance our comprehension of dust production within the CGM.

\subsection{Excess Matter Density in Inner Halos }
In Figure \ref{fig:sigma_compare}, we observe evidence of an excess of matter density in the innermost regions of massive (sub)halos at $r_{\rm{p}}<70h^{-1}$kpc, reaching significance levels of $2.8\sigma$ compared to the DMO model of Planck cosmology and $3.6\sigma$ with the \texttt{TNG300-1} hydrodynamic model. Our result is consistent with the previous findings that the probability of strong lensing events is higher than the $\Lambda$CDM prediction, as noted by \citet{2020Sci...369.1347M} using 11 galaxy clusters, and corroborated by subsequent studies such as \citet{2022A&A...668A.188M} and \citet{2023A&A...678L...2M}, which compared against hydrodynamic simulations. Additionally, \citet{2021PhRvD.104j3031Y} argued that the self-interacting dark matter (SIDM) model \citep{2000PhRvL..84.3760S}, which produces a steeper density profile than $\Lambda$CDM, could account for these strong lensing results. Our measurements also reveal a heightened density in the innermost regions of (sub)halos compared to the DMO model of $\Lambda$CDM. By utilizing 348,938 LRGs as lens galaxies, much larger than the sample of 11 galaxy clusters used in \citet{2020Sci...369.1347M}, we reduce the likelihood that cluster-to-cluster variation could explain these results. As depicted in Figure \ref{fig:sigma_compare}, the incorporation of current galaxy formation models only exacerbates the tension, given that strong baryon feedback models are typically employed to prevent the formation of excessively massive galaxies. Hence, our findings suggest that our understanding of either the nature of dark matter or the processes governing galaxy formation remains incomplete. 

The magnification signal can be more precisely measured in future surveys, providing improved constraints on the inner density profiles of (sub)halos. Furthermore, the velocity dispersion in the inner halos, measured through redshift space distortion \citep[RSD;][]{2012ApJ...758...50L}, is anticipated to be sufficiently accurate in ongoing or next-generation redshift surveys to detect such deviations.   

\section{Conclusion}\label{sec:conclusion}
In this study, we measure the magnification of DECaLS background galaxies lensed by CMASS galaxies. We compare magnification measurements in three different bands ($grz$) to constrain the dust attenuation in the CGM of CMASS galaxies. Subsequently, we derive the true magnification signal after correcting for dust attenuation. Finally, we compare our measurements with model predictions from various cosmological and galaxy formation models, leveraging high-resolution simulations, the precise galaxy-halo connection of CMASS from PAC, and the accurate ray-tracing algorithm \texttt{P3MLens}. Our main findings can be summarized as follows:
\begin{itemize}
    \item We introduce a novel method to measure the magnification signal around lens galaxies by focusing on the change in total flux density, $\delta M$, of source galaxies. This approach is more robust against imperfect deblending issues compared to measuring changes in number density and mean flux. We then convert $\delta M$ to the lens parameter \(\mu\) by establishing the \(\delta \mu-\delta M\) relation using a deeper photometric survey.
    \item We achieve robust magnification measurements across a broad range of physical scales, from $0.016h^{-1}{\rm{Mpc}}$ to $10h^{-1}{\rm{Mpc}}$, with enhanced accuracy observed at smaller scales. These results remain consistent across source samples in different bands and with varying magnitude limits.
    \item We successfully measure the dust attenuation in the CGM of CMASS galaxies by comparing the magnification measurements in $grz$ bands. Our analysis reveals a steep dust attenuation curve, suggesting the potential presence of small dust grains within the CGM of massive galaxies. Additionally, we observe a decrease in dust attenuation with distance, indicating a possible correlation between the dust distribution and the matter distribution.
    \item At large scales ($r_{\rm{p}}>70\,h^{-1}$kpc), our magnification measurement is in good agreement with the predictions from the DMO simulations in both WAMP9 and Planck18 cosmologies. Assuming a linear correlation between $\Sigma_{\mu}$ and $S_8$, we obtain a {\it{rough estimation}} of $S_8=0.816\pm0.024$. Additionally, we observe agreement between predictions from the hydrodynamic model \texttt{TNG300-1} and those from DMO models at large scales.
    \item At small scales ($r_{\rm{p}}<70\,h^{-1}$kpc), we observe that our magnification measurements exceed the predictions from the DMO model of Planck18 cosmology, exhibiting a deviation of $2.8\sigma$. Incorporating the \texttt{TNG300-1} galaxy formation models further exacerbates this tension to $3.6\sigma$. These results point towards a higher matter density in the inner regions of massive (sub)halos, underscoring our incomplete understanding of either the nature of dark matter or the processes governing galaxy formation.
\end{itemize}

Our work reveals that magnification can complement shear in lensing measurements, particularly at small scales where it can be more accurately measured. With upcoming next-generation multi-band photometric surveys and spectroscopic surveys such as the Dark Energy Spectroscopic Instrument (DESI; \cite{2016arXiv161100036D}), the Subaru Prime Focus Spectrograph (PFS; \cite{2014PASJ...66R...1T}), the Legacy Survey of Space and Time (LSST; \cite{2019ApJ...873..111I}), Euclid (\cite{2011arXiv1110.3193L}), the Chinese Space Station Optical Survey (CSS-OS; \cite{2019ApJ...883..203G}), and surveys conducted with the Nancy Grace Roman Space Telescope (Roman; \cite{2015arXiv150303757S}), lensing magnification holds promise for precise measurements of dust attenuation and matter distribution in the universe. These measurements have the potential to advance our understanding of the nature of dark matter and the physics of galaxy formation.

%% Also note that the akcnowlodgment environment does not support long amounts of text. If you have a lot of people and institutions to acknowledge, do not use this command. Instead, create a new \section{Acknowledgments}.
\section*{Acknowledgments}
K.X. thanks Carlos Frenk, Shaun Cole, Sownak Bose, Willem Elbers and Qiuhan He for useful discussions. The work is supported by NSFC (12133006, 11890691), National Key R\&D Program of China (2023YFA1607800, 2023YFA1607801), grant No. CMS-CSST-2021-A03, and by 111 project No. B20019. We gratefully acknowledge the support of the Key Laboratory for Particle Physics, Astrophysics and Cosmology, Ministry of Education. This work made use of the Gravity Supercomputer at the Department of Astronomy, Shanghai Jiao Tong University. 

Funding for SDSS-III has been provided by the Alfred P. Sloan
Foundation, the Participating Institutions, the National Science
Foundation, and the US Department of Energy Office of Science. The SDSS-III web site is http://www.sdss3.org/. 

The Legacy Surveys consist of three individual and complementary projects: the Dark Energy Camera Legacy Survey (DECaLS; Proposal ID \#2014B-0404; PIs: David Schlegel and Arjun Dey), the Beijing-Arizona Sky Survey (BASS; NOAO Prop. ID \#2015A-0801; PIs: Zhou Xu and Xiaohui Fan), and the Mayall z-band Legacy Survey (MzLS; Prop. ID \#2016A-0453; PI: Arjun Dey).

\bibliography{sample631}{}
\bibliographystyle{aasjournal}



\end{document}